\documentclass[twocolumn]{aastex62}

\graphicspath{{./}{figures/}}

\received{January 1, 2018}
\revised{January 7, 2018}
\accepted{\today}
\submitjournal{ApJ}

\shorttitle{Effects of Tellurics in Iodine RVs}
\shortauthors{Wang et al.}

\usepackage{xcolor,fontawesome}
\definecolor{twitterblue}{RGB}{64,153,255}
\newcommand{\twitter}[1]{\href{https://twitter.com/#1}{\textcolor{twitterblue}{\faTwitter}\,\tt \textcolor{twitterblue}{@#1}}}

\usepackage{amsmath,amssymb}
\usepackage{xspace}
\usepackage{graphicx}
\usepackage{epstopdf}
\usepackage{graphicx}
\usepackage{epsfig}
\usepackage{natbib}
\usepackage{verbatim}
\usepackage{morefloats} 

\usepackage[caption=false]{subfig}
\usepackage{float}

\usepackage{CJK}

\def\beq{\begin{equation}}
\def\eeq{\end{equation}}
\def\bcm{}

\def\kepler{{\it Kepler}}
\def\keck{Keck/HIRES}

\def\leq{\leqslant}

\newcommand{\rev}[1]{\textcolor{black}{#1}}

\begin{document}

\begin{CJK*}{UTF8}{gbsn}

\title
{
The Effects of Telluric Contamination in Iodine Calibrated Precise Radial Velocities
\footnote{Based on observations obtained at the Keck Observatory,
which is operated by the University of California. The Keck
Observatory was made possible by the generous financial support of the
W. M. Keck Foundation.}
}

\correspondingauthor{Sharon Xuesong Wang}
\email{sharonw@carnegiescience.edu}

\author[0000-0002-6937-9034]{Sharon Xuesong Wang (王雪凇)}
\affiliation{The Observatories of the Carnegie Institution of Washington, 813 Santa Barbara Street, Pasadena, CA 91101, USA \\ \twitter{sharonxuesong}}

\author{Jason T.\ Wright}
\affiliation{Department of Astronomy and Astrophysics, 525 Davey
  Laboratory, The Pennsylvania State University, University Park, PA
  16802, USA}
\affiliation{Center for Exoplanets and Habitable Worlds, 525 Davey
  Laboratory, The Pennsylvania State University, University Park, PA
  16802, USA}

\author{Chad Bender} 
\affiliation{Steward Observatory, University of Arizona, Tucson, AZ 85721, USA}

\author{Andrew W. Howard} 
\affiliation{Department of Astronomy, California Institute of Technology, Pasadena, CA 91125, USA}

\author{Howard Isaacson} 
\affiliation{Department of Astronomy, University of California, Berkeley, CA 94720, USA}

\author[0000-0002-0385-2183]{Mark Veyette} 
\affiliation{Department of Astronomy \& Institute for Astrophysical Research, Boston University, 725 Commonwealth Avenue, Boston, MA 02215, USA}

\author[0000-0002-0638-8822]{Philip S. Muirhead} 
\affiliation{Department of Astronomy \& Institute for Astrophysical Research, Boston University, 725 Commonwealth Avenue, Boston, MA 02215, USA}



\begin{abstract}

We characterized the effects of telluric absorption lines on the radial velocity (RV) precision of stellar spectra taken through an iodine cell. To isolate the effects induced by telluric contamination from other stellar, instrumental, or numerical systematic RV noise, we extracted RVs from simulated iodine calibrated spectra of three RV standard stars regularly observed by \keck. We add in water absorption lines according to measured precipitable water vapor (PWV) contents over a one-year period. We concluded that telluric contamination introduces additional RV noise and spurious periodic signals on the level of 10-20~cm/s, consistent with similar previous studies. Our findings show that forward modeling the telluric lines effectively recovers the RV precision and accuracy, with no prior knowledge of the PWV needed. Such a recovery is less effective when the water absorption lines are relatively deep in the stellar template used in the forward modeling. Overall, telluric contamination plays an insignificant role for typical iodine-calibrated RV programs aiming at $\sim$1--2~m/s, but we recommend adding modeling of telluric lines and taking stellar template observations on nights with low humidity for programs aiming to achieve sub-m/s precision.

\end{abstract}

\keywords{techniques: radial velocities --- 
techniques: spectroscopic --- atmospheric effects}




\section{Introduction}\label{sec:intro}

The first exoplanets around main-sequence stars were discovered with the
radial velocity (RV) method, where precise Doppler spectroscopy
measures the wavelength shift of the host stars induced by the
gravitational pull of the planets \citep{1988ApJ...331..902C,
  1989Natur.339...38L, 1993ApJ...413..339H, 1995Natur.378..355M,
  1996ApJ...464L.153B}. Since then, the RV method has been used to discover
hundreds of planetary systems (see exoplanets.org; \citealt{eod2014})
and contributed to numerous confirmations and characterizations of
exoplanets discovered with the transit method (e.g., for
\kepler\ follow-up observations; \citealt{Marcy2014}).

The current best RV precision being readily achieved is around 1~m/s \citep{eprv2015} \rev{and occasionally for quiet and bright stars, below 1~m/s \citep[e.g.,][]{pepe2011},}
mostly commonly attainable via two wavelength calibration methods in the optical band:
ThAr lamp emission line calibration (e.g., ELODIE and HARPS;
\citealt{elodie, harps-s}; $\sim$400--690~nm) and iodine cell
absorption line calibration (e.g., Keck/HIRES and Magellan/PFS;
\citealt{1994SPIE.2198..362V, butler1996, 2010SPIE.7735E..53C}; $\sim$500--620~nm). Over 20 new precise RV spectrographs are being built and commissioned as of 2018, and most of them aim at an RV precision of better than 1~m/s or even 10~cm/s \citep{eprv3}. The
major obstacles for achieving a higher RV precision include stellar
activity induced RV signals, instrumental effects, telluric
contamination, and limitation in data analysis \citep{eprv2015}.

Traditionally, telluric contamination has not been considered as problematic
for precise RV in the optical. It is certainly, however, a severe source of
spectral contamination and a bottleneck for achieving higher RV
precision in the near infra-red (NIR) region (e.g.,
\citealt{2010ApJ...713..410B}), where a large number of deep water and
methane lines reside. However, only a small wavelength
range exists in the optical that has deep telluric lines, and typically, such
regions are simply omitted for the purpose of precise RV analysis,
either by giving them zero weights in the cross-correlation masks (for
ThAr calibrated spectra, e.g., \citealt{2002A&A...388..632P}) or
flagging them as bad pixels (for iodine-calibrated spectra, e.g., for
Keck/HIRES).

The works by \cite{artigau2014} and \cite{cunha2014} have
characterized and mitigated the effects of telluric contamination in
the precise RV data taken by the ThAr-calibrated HAPRS.
\cite{cunha2014} focuses on the issues with ``micro-telluric" lines
(shallow telluric absorption lines with $<1$--3\% depths), which are
considered in the context of precise RV for the first time. \cite{cunha2014} fit and then divided
out the telluric lines in the observed spectra using synthetic
telluric spectra generated by the LBLRTM package (Line-By-Line
Radiative Transfer Model, \citealt{lblrtm}; with line lists from
HIgh-resolution TRANsmission molecular absorption database, or HITRAN,
\citealt{hitran2013}) and TAPAS \citep{tapas}, which is a more
user-friendly, though less flexible, package using LBLRTM. They
concluded that the micro-tellurics have an impact (defined as the root mean square [RMS] of the
difference between RVs before and after micro-telluric removal) of
$\sim$10--20 cm/s for G stars observed with low to moderate air masses,
but the impact can be substantial in some cases, up to $\sim$0.5--1
m/s.

\cite{artigau2014} uses principal component analysis (PCA) to
empirically correct for telluric lines in HAPRS data (both
micro-tellurics and the deep lines in the $\sim$630~nm region). 
Combining PCA with rejection masking, they have reduced the RV RMS by
$\sim$20~cm/s (and more significantly for the $\sim$630~nm
region). More recently, \cite{2016AAS...22713719S} characterized the
effects of telluric contamination and effectiveness of some typical
remedies (masking and modeling) for emission line-calibarated spectra
for the optical, broad optical (300--900~nm), and NIR. Their conclusion
for the optical region is similar to the results in \cite{artigau2014}
and \cite{cunha2014}, but in the NIR, even if all of the telluric lines are modeled and subtracted to the 1\% level, the residuals would still cause 0.4--1.5 m/s RV errors for M and K dwarfs.

This paper characterizes and corrects for the adverse effects of telluric contamination\footnote{Because of the wavelength range of our work ($\sim$500--620~nm), we focused on the impact of water absorption, with some consideration of oxygen lines. We simply refer to the water and oxygen absorption as the ``telluric absorption" throughout this paper, although this term normally means more than just absorption lines from these two molecular species.} under the context of iodine-calibrated precise RVs. In particular, we focused on quantifying the RV errors and systematics induced by tellurics and the effectiveness of mitigation methods, both in idealized cases and under realistic scenarios. We used simulated spectra in order to isolate the effects of tellurics, and we chose forward modeling as the method for mitigating tellurics instead of dividing the telluric lines out (which is mathematically incorrect because convolution, i.e., line broadening by the spectrograph, is not distributive over multiplication; \citealt{muirhead2011}). 

The paper is structured as follows: we describe our methodology in Section~\ref{sec:method} and detail our findings in Section~\ref{sec:results}; and we comment on results with real iodine-calibrated RV observations in Section~\ref{sec:realobs} and summarize our conclusions and recommendations in Section~\ref{sec:summary}.

\section{Methodology}\label{sec:method}

To quantify the impacts of micro-telluric absorption features (often referred to simply as ``tellurics'' below) and isolate their effects, we performed an end-to-end simulation, from the simulated iodine-calibrated precise 1-D RV spectra to the data analysis process on RV standard stars. Real iodine-calibrated RV data have systematic errors stemming from errors in the deconvolved stellar template \citep{thesis} and other unknown sources (typically referred to as the instrumental RV ``jitter").

We used the \keck\ spectral format for our study because \keck\ is one of the most widely used iodine-calibrated spectrometers with high RV precision, and it has long observing baselines on many RV standard stars. Notably, the results in this work could be of general application despite us selecting one specific instrument, because most iodine-calibrated PRV instruments are similar in terms of resolution, sampling factor on the CCD, spectral grasp, and signal-to-noise (SNR) ratios for their observations.

Throughout this study, we used the atmospheric conditions and weather data of the Kitt Peak National Observatory as the input into our simulated spectra. This is because Mauna Kea is significantly drier than most observatories, and thus, the telluric lines in the iodine region (mostly water lines) are shallower. This does not alter the main conclusion of our study, which is that tellurics have a negligible impact on the RV precision for typical iodine-calibrated spectra, aiming at 1--2~m/s.

\subsection{Choice of Stars and Synthetic Stellar Spectra\label{sec:stars}}

We performed simulations with three stars: HD 185144 ($\sigma$ Dra), HD 10700 ($\tau$ Ceti), and HD 95735 (GJ 411), which are among the benchmark RV standard stars because they are bright, quiet, and not hosts of known planets.\footnote{There is a claimed planetary system around HD 10700 by \cite{tuomi2013} and \cite{feng2017}, but the amplitudes of the claimed planets are very small compared with the overall RV RMS of the star, and thus, they do not affect our experiments and general conclusion with the real \keck\ observation in Section~\ref{sec:realobs}. The existence of planet candidates certainly does not affect our simulations, which all have an input of RV $=$ 0~m/s.} They exhibit the smallest RV variation on both short timescales of days and also long timescales of years. The RV data on standard stars are often good diagnostic tools for identifying RV systematics. We simulated the spectra of these three stars with the same SNR and observing cadence as the real \keck\ data.

HD 185144 is a G9V star (SIMBAD), and it is among the most frequently observed star with \keck. It has 712 \keck\ observations as of early 2016,\footnote{\keck\ has continued to add additional observations, but the increase in sample size is not important for our purposes here.} with RV RMS $=$ 2.2 m/s \citep{butler2017}, and it has a relatively small span in barycentric velocity (often referred to as the barycentric velocity correction, or BC), $[-4.7,\ 4.7]$ km/s, because of its proximity to the northern ecliptic pole. RV systematics tend to correlate strongly with BC, because the largest Doppler signal in the stellar spectrum is the BC component, and thus, BC dictates both the position of the stellar spectrum on the CCD and which iodine lines it interacts with. Due to HD 185144's small BC span, its RV data sample the BC space more densely than most stars, and therefore, in general, RV systematics would show up most clearly in an RV vs.\ BC plot of HD 185144. 

HD 10700 is a G8.5V star, and it has 623 observations as of early 2016, with RMS $=$ 2.4 m/s. Its BC span is $[-27.8,\ 26.8]$ km/s, one of the largest due to its proximity to the ecliptic equator. HD 95735 is a M2V star, and it has 243 \keck\ observations as of early 2017. Its RV RMS is 3.3 m/s, with a BC span of $[-26.4,\ 27.0]$ km/s.

To construct simulated spectra, we used synthetic stellar spectra instead of observationally derived stellar spectra, because the observationally derived stellar spectra contain telluric absorption lines. The synthetic spectra of HD 185144 and HD 10700 were generated using Spectroscopy Made Easy (SME; \citealt{valentipiskunov1996,valentifischer2005}; the SME spectra are provided by Dr.\ Jason Curtis). They were generated using the best-fit stellar parameters estimated by SME using selected spectral windows that are sensitive to key stellar parameters, such as effective temperature and gravity \citep{valentifischer2005}. The original SME output synthetic spectra are continuum normalized and have a resolution of $R \sim 1,000,000$, convolved with a single gaussian spectral point spread function. The stellar parameters for HD 185144 are ${\rm T_{eff}} = 5246 K$, $\log{g}=4.55$, and $[M/H] = -0.16$. The stellar parameters for HD 10700 are ${\rm T_{eff}} = 5283 K$, $\log{g}=4.59$, and $[M/H] = -0.36$. The best-fit $v\sin{i}$ values and the default macro turbulence parameters for these two stars would yield spectral lines that are too broad compared to their $R \sim 300,000$ deconvolved stellar spectral templates (DSSTs, referred to as ``stellar templates" below), derived observationally from \keck\ data, so we turned off rotational and macro turbulence broadening. The output SME spectra have lines that are slightly narrower than their corresponding stellar templates, but we did not convolve the SME spectra to match the line widths in stellar templates, to avoid introducing additional numerical errors. This did not affect our study since we simulated and fit observed spectra with much lower resolutions at $R\sim 70,000$ and $R\sim 120,000$.
  
The HD 95735 synthetic spectrum is from \cite{coelho2014}, provided by Dr.\ Paula Coelho. It has a spectral resolution of $R = 300,000$ and the input stellar parameters ${\rm T_{eff}} = 3600 K$, $\log{g}=4.5$, and $[M/H] = -0.5$, which were chosen based on visual inspection to find the best match among the synthetic model spectra grids and HD 95735's Keck stellar template, in terms of line density and line depths. These stellar parameters are also consistent with the various measurements reported on SIMBAD. At such high resolution, the synthetic spectrum does not match the stellar template in most lines due to the limitation of the molecular line data in the optical at such a high resolution.

We also generated synthetic spectra based on the BT-Settl atmospheric models \citep{allard2012a,allard2012b,baraffe2015} to see it would better match the observed spectrum. We used version 15.5 of the PHOENIX stellar atmosphere modeling code to produce synthetic spectra. The PHEONIX and Coelho models look very similar. They both have similar variance in flux with the deconvolved observed spectrum, indicating similar Doppler content in the stellar lines, which was sufficient for the purpose of our study.

\subsection{Generating Synthetic Telluric Spectra}\label{sec:syntelluric}

\begin{figure}
\includegraphics[scale=0.35]{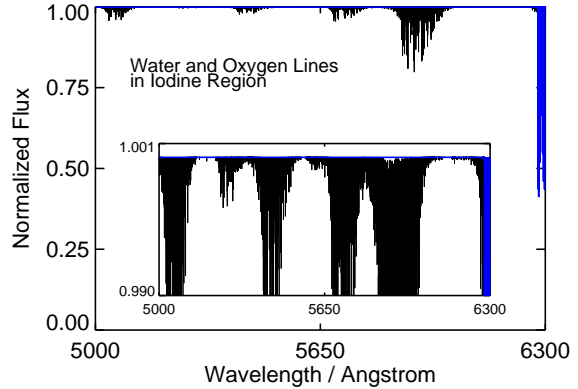} 
\caption{Telluric lines in the iodine region are mostly shallow water
lines, with some moderately deep water lines near 5900\AA\ and very
deep oxygen lines near 6300\AA. Blue lines are oxygen lines. The PWV in the plotted spectrum is 4.0 mm, typical for Kitt Peak but slightly humid for Mauna Kea. The atmospheric profile used is tropical (appropriate for Mauna Kea) with airmass 1.0. The spectrum is generated using TERRASPEC \citep{Bender2012} at a fully resolved resolution. The insert plot is showing the pervasiveness of micro-telluric lines, i.e., lines with $\leq$1\%--3\% depths.
\label{fig:telluric}}
\end{figure}

We used two software packages, TERRASPEC \citep{Bender2012} and TAPAS \citep{tapas}, to generate synthetic telluric spectra, and verified that they provide the same answer. Both packages were based on HITRAN line list \citep{hitran2013} and the LBLRTM package \citep{lblrtm}. We used the ``mid-latitude summer'' atmospheric profile as the input for atmospheric layers, which is appropriate for Kitt Peak. Since the telluric lines in the iodine region are mostly water and oxygen lines, we separately generated synthetic spectra for water and oxygen, without accounting for Rayleigh scattering. The output synthetic telluric spectra have a spectral resolution higher than 1,000,000. An example of generated synthetic spectrum is shown in Figure~\ref{fig:telluric}.

We changed line depths of water or oxygen lines by scaling them with a power law. The water line depths varied because of changes in the amount of precipitable water vapor (PWV) in the atmosphere and because of airmass change, and we computed the scaled flux as $f = f_0^{{\rm PWV \times airmass}}$, where $f_0$ is the absorption spectrum at airmass = 1 and PWV = 1~mm. This scaling worked well for water lines because they are mostly optically thin in the iodine region. As a result, the transmission follows $f = e^{-\tau}$, and the optical depth $\tau$ is proportional to the amount of water along the line of sight, i.e., PWV$\times$airmass, in a simple plane parallel atmospheric model.  

The oxygen line depths varied mostly because of airmass changes since oxygen is a very stable and well mixed constituent of the atmosphere. We also scaled the oxygen lines using airmass following the above formula; although the oxygen lines are not all optically thin, this scaling works well for all unsaturated lines. We have verified that this simple power law scaling is good enough to capture the line changes due to PWV and airmass by comparing the scaled spectra with models generated by TERRASPEC and TAPAS, but notably, modeling oxygen lines in real observations is a much more challenging task (see, e.g., \citealt{figueira2012}). We used such simpling scaling to vary the telluric line depths because running TERRASPEC or TAPAS is time consuming, and it would be computationally inefficient to fold the atmospheric synthesizing code into the forward modeling code for extracting RVs from spectra.

\begin{figure}
\includegraphics[scale=0.35]{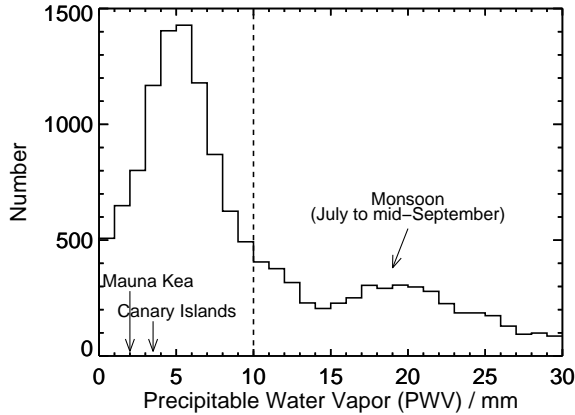} 
\caption{Histogram of amount of PWV in mm at the Kitt Peak National Observatories from SuomiMet \citep{suominet2000}. Data were measured in half-hour intervals during the entire year of 2016. PWV values used in our simulations were randomly drawn from this distribution, but limited to PWV $< 10$ mm. \rev{For comparison, the median PWV of Mauna Kea and Canary Islands are also marked.} See Section~\ref{sec:syntelluric} for more details.
\label{fig:pwvhist}}
\end{figure}

Since our simulated spectra are based on real \keck\ observations, we adopted the airmass values of these observations in our simulation to scale the water and oxygen lines. For the PWV values, we used the SuomiNet \citep{suominet2000} water monitoring data from their station at Kitt Peak, which registers PWV values throughout the year at a 30-minute cadence. We used the 2016 PWV data\footnote{\url{http://www.suominet.ucar.edu/data/staYrHr/KITTnrt_2016.plot}} and draw randomly from the distribution of valid PWV values but restricted it to values less than 10 mm since nights with larger PWV values are often associated with bad weather conditions that are not optimal for astronomical observations. Ten mm is essentially a conservative cut, but it does not alter the conclusion of our study. Figure~\ref{fig:pwvhist} illustrates the PWV distribution of Kitt Peak in 2016. \rev{For comparison, the median PWV values of Mauna Kea and the Canary Islands are also marked, which are 2~mm and 3.5~mm, respectively. These two sites are significantly dryer than Kitt Peak, having 90\% or more of the nights with PWV$<$10~mm (see \citealt{sarazin2002} and \citealt{garcia2010} for Mauna Kea and \citealt{castro-almazan2016} for the Canary Islands).}

\subsection{Simulating Iodine-calibrated Stellar Spectra}\label{sec:simspec}

For the three stars chosen in this study, we simulated two sets of spectra for each real \keck\ observation: one set with spectral resolution R $\sim$ 70,000 (70k; e.g., typical \keck\ resolution), and the other set with R $\sim$ 120,000 (120k; e.g., the resolution of Magellan/PFS2; \citealt{pfs2006,pfs2010}). The \keck\ CCD has a sampling factor of $\sim$3 pixels per resolution element when using the C2 or B5 decker (with R $\sim$ 55,000, though this is normally higher and reaching up to 70,000 because of the good seeing at Mauna Kea).  For each spectral resolution, we simulated spectra with and without photon noise. The simulated spectra with photon noise have the same SNR per pixel as its corresponding \keck\ observation, typically SNR $\sim$ 200 per pixel. All simulated spectra have one version free of telluric lines, and another one with telluric lines added, generated using the randomly drawn PWV value and the airmass value from the corresponding \keck\ observation. No {\it intrinsic} stellar velocity variation exists in the simulated spectra --- i.e., the spectra are only Doppler shifted according to the barycentric velocity of the Earth at the time of observation.

Below is our recipe for generating a simulated iodine-calibrated spectrum:

\begin{enumerate}
  \item Shift the synthetic stellar spectrum to have the same barycentric velocity
    as the star at the epoch of the corresponding \keck\ observation.
  \item Generate a continuous wavelength solution for each echelle
      order based on the best-fit wavelength solution for the corresponding \keck\
      spectrum. 
  \item Re-sample the synthetic spectrum and the iodine atlas onto a
    wavelength grid that is four times finer than the wavelength
    solution grid.\footnote{Ideally, one would multiply the two spectra at a higher sampling and then re-sample to a coarser grid. However, we chose to sample down the synthetic spectrum here before multiplication because the input stellar template being used to forward model these simulated spectra was down sampled as such to match the real \keck\ templates. This way, our synthesizing process for making simulated spectra was numerically identical to the synthesizing process used in the forward modeling code, which eliminated potential additional numerical errors. } We chose to re-sample at this sampling factor to match the spectral synthesis procedure in the forward modeling code. Then multiply the stellar spectrum with the iodine atlas. 
  \item Add telluric contamination by multiplying the spectrum produced in the Step 3 with a telluric absorption spectrum scaled using the appropriate airmass and the randomly drawn PWV value.
  \item Convolve each order of the multiplied spectrum with
    a single Gaussian line spread function (LSF).\footnote{The LSF model for real \keck\ observations is a multi-Gaussian model, with a central Gaussian with fixed width and satellite Gaussian pairs with fixed positions and width but varying heights \citep{valenti1995}. We also performed simulations using a multi-Gaussian LSF input and found the same conclusion. Here, we only present results from simulations using single Gaussian LSF input for simplicity.} We used a Gaussian LSF with $\sigma=1.7$ pixels for R = 70k and $\sigma=1.0$ pixel for R = 120k. Then we re-sampled the convolved spectrum down to the observed wavelength pixel grid using a flux conservative algorithm.
  \item Fit for the blaze function for each order of the corresponding \keck\
    spectrum with a third-order polynomial using the top 1\% of the flux,
    and add this fitted blaze function to the convolved spectrum. This ensures that the simulated spectrum has the same photon counts as the corresponding \keck\ spectrum.
  \item Add photon noise according to Poisson statistics, when appropriate. For the pair of simulated spectra with and without telluric lines, we added the exact same photon noise in each pixel for each pair to ensure that the change in reported RVs was purely due to the added tellurics instead of variance in the randomly generated photon noise.
\end{enumerate}

Extracting RVs from iodine-calibrated spectra is done via forward modeling (\citealt{butler1996}; see the next section), which requires a stellar template (or DSST; as mentioned in Section~\ref{sec:stars}). We simply converted the synthetic stellar spectra into the stellar template format needed by the Doppler code (with matching spectral sampling and chunking as the real \keck\ stellar template for each star). We shifted the wavelengths of synthetic stellar spectra to be at the same wavelengths with the corresponding \keck\ observations that generated the stellar templates, i.e., accounting for each star's systemic velocity and the barycentric velocity of the Earth at the epoch of the stellar template observation. As a result, just as for real observations, all reported RVs were the stellar velocities {\it relative to} the star's velocity at the epoch of its template observation. We added telluric absorption lines into the stellar templates since, in reality, the stellar template is derived from on-sky observations. Eliminating telluric absorption lines from DSST is not a trivial process given the shallow line depths of water lines and the fact that they are heavily blended with the stellar lines. We have experimented with cleaning telluric lines for DSST in real observations, but, given the resolution, sampling, and SNR of the real observations, the extra uncertainties induced by this ``cleaning'' process introduced more errors than the errors induced by the tellurics \citep{thesis}. Better algorithms for generating stellar templates may yield a different or better result, but are beyond the scope of this paper. Therefore, for our simulations, we did not use telluric-free stellar templates. We used two versions of stellar templates: PWV = 1 mm (simulating template observations taken on a very dry night) and PWV = 5 mm (taken on a typical night), both with airmass = 1.0.

\subsection{The Doppler Code for Extracting RVs}\label{sec:dopcode}

We used the California Planet Survey (CPS) Doppler code to extract precise RVs from the simulated spectra. The code uses forward modeling to fit the iodine-calibrated stellar spectra, following the algorithm outlined in \cite{butler1996}. Section~2 of \cite{thesis} has a more detailed documentation on the code, and some of its elements are described in \cite{2006ApJ...647..600J}, \cite{2009ApJ...696...75H,2011ApJ...726...73H}, and \cite{2011ApJS..197...26J}. Our copy of the code was kindly provided by John A.\ Johnson and the CPS group in 2013. 

The Doppler code takes the DSST and the iodine atlas as input model spectra, and fits the observed spectra to solve for the wavelength solution and Doppler shift. Each observed spectrum at a given epoch is divided into 80 pixel chunks (about 2 \AA) and fitted independently using the Levenberg-Marquardt least-$\chi^2$ algorithm. Later, the RVs reported from all chunks are combined through an outlier rejection and weighted averaging process to yield the final RV for the given epoch. The free parameters in our fitting for a telluric-free spectral chunk include $\lambda_0$, the wavelength for the first pixel of this chunk (\AA); $\delta \lambda$, the linear wavelength dispersion (\AA\ per pixel); the width of a single Gaussian LSF, $\sigma$ (pixel); and the Doppler shift, $z$. Although we knew the exact width for the LSF in our simulated spectra, we chose to float $\sigma$ as a free parameter because, otherwise, the reported RVs contain additional systematics, probably induced by numerical errors in the making of the simulated spectra. For the simulated noise-free spectra, we used a quadratic wavelength solution in the fitting instead of a linear one, which yields a significantly higher RV precision (20--30 cm/s vs. $<$ 10 cm/s). For simulated spectra with noise added, we did not invoke quadratic wavelength solution because it would have induced additional numerical errors for spectra with more realistic SNRs.

\begin{figure}
\includegraphics[scale=0.35]{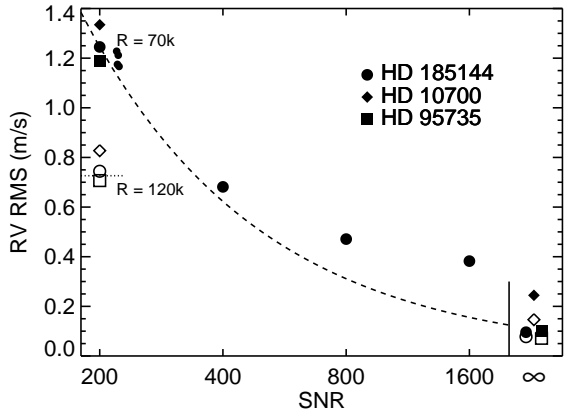} 
\caption{\rev{RV RMS vs.\ SNR per pixel in the simulated \textbf{telluric-free} spectra for each star to demonstrate the precision of the adopted Doppler pipeline}. Filled symbols are for the simulated spectra with resolution $R=70,000$ or 70k, and hollow symbols are for the higher resolution spectra ($R=120,000$ or 120k). The dashed line is the predicted RV precision scaling relation assuming RMS $\propto$ SNR$^{-1}$ \citep[e.g.,][]{butler1996}. The dotted line marks the scaled precision at $R=120$k assuming RMS $\propto R^{-1}$. Both scaling relations are normalized at SNR = 200, RMS = 1.25 m/s for HD 185144 with resolution $R=70$k. The small dots (shifted horizontally a bit off SNR = 200 for visualization) are RMS for other sets of simulated spectra of HD 185144 with SNR = 200 and $R=70$k with different photon noise added for each set. The set of points to the right of the vertical line is RV RMS from noise-free spectra (labeled with $\infty$ for SNR). See \ref{sec:dopcode} for more information.
\label{fig:snrprecision}}
\end{figure}

Figure~\ref{fig:snrprecision} illustrates the precision of the Doppler code for spectra with different SNR and resolution (all from telluric-free simulations). The filled black dots are RMS values of the RVs for simulated observations of HD 185144 at different SNRs with a resolution of $R$ = 70k. They roughly follow RMS $\propto$ SNR$^{-1}$ \citep[e.g., ]{butler1996}, but have lower precision than predicted at high SNR because of inaccuracies in the wavelength solution for each chunk -- the model used is only linear in pixel space and does not have higher-order terms. The precision is recovered once a quadratic wavelength solution for each chunk is implemented, for example, for the simulations without photon noise added, plotted in the lower right corner labeled with $\infty$ for SNR. The high SNR results are not realistic; rather, they are just an estimate of the code's precision -- CCDs very often enter the non-linear regime or saturate near and beyond SNR = 400. Figure~\ref{fig:snrprecision} shows that the code performs well since the RV RMS or precision scales as expected with SNR and spectral resolution, and it is capable of producing sub-m/s RV precision for spectra with high enough resolution and SNR. The RV RMS values reported for the noise-free cases represent a floor of systematic RV error from the forward modeling code itself.

When fitting simulated spectra injected with telluric absorption lines, we either ignored the telluric lines (i.e., did not incoporate any special treatment for them, such as masking them out), to assess the impacts of tellurics to compare the results with telluric-free spectra, or incorporated a component of telluric spectrum in the forward modeling to assess how effective modeling mitigates the effects of tellurics. When including a telluric component in the forward modeling, an additional free parameter for the PWV was introduced to fit for water absorption lines by scaling the input water spectrum in power law with PWV $\times$ airmass (see Section~\ref{sec:syntelluric}). The input model water spectrum was the same one used for synthesizing the simulated spectra, with PWV = 1~mm. Because most spectral chunks have neglegible amount of water absorption, only chunks with more than 0.1\% water absorption (when PWV = 1~mm) were fitted with a telluric component. We did not attempt to fit or offset the telluric lines in the stellar templates (see more in the last paragraph of Section~\ref{sec:syntelluric}).

When the code fit for the telluric component, it introduced an additional ``pass'' in the least-$\chi^2$ fitting procedure: it first fit for the PWV in the first pass, and then fixed the PWV to the best-fit value in the subsequent passes where it fit other parameters, including the Doppler shift. The fixed best-fit PWV value was the numerical median of all fitted values from chunks with more than 0.1\% water absorption.

In principle, we could have also fit the oxygen lines near 6300\AA. We chose not to do so in this paper, because the choice would not alter our conclusion. For real observations, it is extremely hard to fit the oxygen lines to a high enough precision---it is challenging for the synthetic telluric model to accurately reproduce the shapes of these deep absorption lines. The line shapes closely depend on the conditions of atmospheric layers since oxygen has a significant scale height in the atmosphere. The exact line centers or wavelengths could also change, for example, due to wind \citep{figueira2012}. As a result, oxygen line models would leave a significant residual after fitting and still affect the quality of the RVs out of these spectral chunks. In addition, the wavelength region beyond 6200\AA\ has significantly fewer iodine absorption lines and thus much less information content. The results of both of these effects would give rise to very large RV RMSs for spectral chunks contaminated with oxygen lines, and therefore, these chunks would be rejected when the code computes a final weighted average RV using RVs from all chunks.

\section{Results}\label{sec:results}

In this section, we present the results from the simulations described above. We first quantifed the effects of telluric contamination in iodine-calibrated precise RVs (Section~\ref{sec:impact}) in terms of the amount of added RV RMS (on the level of 10 cm/s), as well as the induced systematics or spurious periodic signals (with amplitudes around 10--20 cm/s at periods of one year and its harmonics; Section~\ref{sec:impactper}). We show that forward modeling the telluric lines effectively mitigate these effects. Then we discuss retrieving the PWV values by forward modeling the water absorption lines in Section~\ref{sec:retrieval}.

\renewcommand{\arraystretch}{1.4} 
\begin{deluxetable*}{rcccccc}
  \centering
\tabletypesize{\scriptsize}
\tablecaption{List of Tests and Results
\label{tab:simulation}}
\tablehead{
  \colhead{Type of Obs.} & \colhead{Spectral} 
  & \colhead{Template} & \colhead{Fit tellurics?} & 
  \colhead{HD 185144 RMS} &
  \colhead{HD 10700 RMS} &
  \colhead{HD 95735 RMS} \\
  \colhead{} & \colhead{Resolution} 
  & \colhead{PWV} & \colhead{} & 
  \colhead{m/s} &
  \colhead{m/s} &
  \colhead{m/s}
}
\startdata
{\bf Noise free\tablenotemark{a}} & & & & & & \\
No tellurics & 70k & \nodata & \nodata & 0.10 & 0.24 &	0.10 \\
With tellurics & 70k & 1 mm & No & 0.13	&	0.29	& 0.14\\
With tellurics & 70k & 1 mm & Yes & 0.11	&	0.22	&	0.12 \\
With tellurics & 70k & 5 mm & Yes & 0.15	&	0.20	&	0.14 \\
No tellurics & 120k & \nodata & \nodata &0.08	&0.15	&0.07 \\
With tellurics & 120k & 1 mm & No & 0.09	&0.18	&0.10\\
With tellurics & 120k & 1 mm & Yes & 0.08	&	0.14	&	0.08 \\
With tellurics & 120k & 5 mm & Yes & 0.09	&	0.15	&	0.09 \\
 \hline
{\bf SNR = 200} & & & & & & \\ 
No tellurics & 70k & \nodata & \nodata & 1.25	&	1.33	&	1.19\\
With tellurics & 70k & 1 mm & No & 1.28	&	1.35	&	1.23 \\
With tellurics & 70k & 1 mm & Yes & 1.24	&	1.29	&	1.23 \\
With tellurics & 70k & 5 mm & Yes & 1.27	&	1.38	&	1.21 \\
No tellurics & 120k & \nodata & \nodata & 0.74	&	0.83	&	0.71 \\
With tellurics & 120k & 1 mm & No & 0.75	&	0.85	&	0.75 \\
With tellurics & 120k & 1 mm & Yes & 0.73	&	0.85	&	0.72 \\
With tellurics & 120k & 5 mm & Yes & 0.75	&	0.89	&	0.73 \\
\hline
{\bf Real Obs.} & & & & & & \\ 
\nodata & $\sim$60--70k & \nodata & No & 2.62 &	3.08 &	3.43 \\ 
\nodata & $\sim$60--70k & \nodata & Yes & 2.71	& 3.13 &	3.34 \\
\nodata & $\sim$60--70k & \rev{cleaned DSSTs} & Yes & 2.75 &	3.66 &	3.23
\enddata
\tablenotetext{a}{With quadratic wavelength solution when forward modeling.}
\end{deluxetable*}

\subsection{Impact of Telluric Contamination on RVs: Increased RV RMS}\label{sec:impact}

Table~\ref{tab:simulation} lists the RV RMS values for the three stars we adopted in various simulations (with or without tellurics, with or without fitting tellurics). Table~\ref{tab:simulation} has three major sections: the top section lists results from simulations using the noise-free spectra, the middle section lists results from simulations using the SNR = 200 spectra, and the last section lists results using real \keck\ observations. The first four columns specify the conditions used in synthesizing the spectra or in extracting the RVs. The first column specifies whether the synthesized spectra used have telluric absorption added. The second column contains the spectral resolution. The third column specifies which stellar template was used when extracting RVs from the synthesized spectra---a telluric free one, or one with water absorption lines with PWV = 1~mm or 5~mm. The fourth column indicates whether a telluric component was fitted or not in the forward modeling process when extracting RVs. The last three columns list the RMS values of RVs for each star in each set of simulation.

Comparing the RV RMS values in row 1 vs.\ row 2 and row 5 vs.\ row 6 of the first two sections in Table~\ref{tab:simulation}, it is evident that telluric absorption adversely affected the RV precision, though the effect is very small. The RV RMS values from telluric-free spectra vs.\ telluric-contaminated spectra only differed by a few cm/s. More precisely, without any treatment, telluric absorption added about 10 cm/s additional RV error in quadrature to the RV RMS (up to 30 cm/s for SNR = 200 cases). This result is similar to the reported values for HARPS in \cite{cunha2014} and \cite{artigau2014}, though smaller in general as we worked with a smaller wavelength coverage with less telluric absorption compared with HARPS (wavelength coverage: 378--691~nm; \citealt{harps-s}). This effect of added RV RMS was more evident in the noise-free simulations, and it definitely existed for situations with realistic SNRs as well (with the added amount of RV RMS consistent with the noise-free ones). However, in practice, the added RV scatter is basically negligible considering that, for real observations, the photon-limited precision is usually much larger than 10 cm/s. This amplitude of 10 cm/s was also on par with the amount of systematic errors from the forward modeling process alone, represented by the RV RMS values for the noise-free telluric-free cases.

The RV RMS values listed in rows 3 and 7 of the first two sections in Table~\ref{tab:simulation} demonstrate the effectiveness of modeling in mitigating telluric contamination. In almost all cases, modeling improved the RV precision, reducing the added RMS by 50\% or more. Modeling of tellurics did not completely recover the lost precision due to errors in modeling the water lines and the existence of telluric lines in the stellar template, as well as systematics introduced into the forward modeling procedure because of the additional parameter of PWV. Again, the noise-free cases saw a more obvious improvement than the SNR = 200 ones (in particular, HD 95735 had a slight increase in RMS when modeling of tellurics was added for the $R$ = 70k and SNR = 200 cases, probably because the effects of tellurics/modeling were buried under noise for this resolution and SNR).

Telluric contamination in the stellar templates would certainly increase the RV RMS as well. With modeling of telluric lines in the observed spectra (and without taking out the tellurics in the stellar templates), simulations using templates with PWV = 5~mm have 5--30 cm/s RV RMS added in quadrature compared with the telluric-free cases. Again, these are not large compared with the typical photon-limited precision, but a general conclusion would be that stellar template observations should be taken on dry nights, if possible. For stellar template spectra with high spectral resolution and high SNR, it is plausible that the telluric lines in the stellar template could be removed by simply dividing them out, although it would not be mathematically correct, but the leftover residual would be smaller than the original water line depths. However, as mentioned in Section~\ref{sec:syntelluric}, we performed this experiment with \keck\ templates and concluded that it does not improve the RV precision. It may require spectra with a higher fidelity (e.g., better known LSF and/or wavelength solution) for this proposed method to be effective.

\begin{figure*}
\includegraphics[scale=0.7]{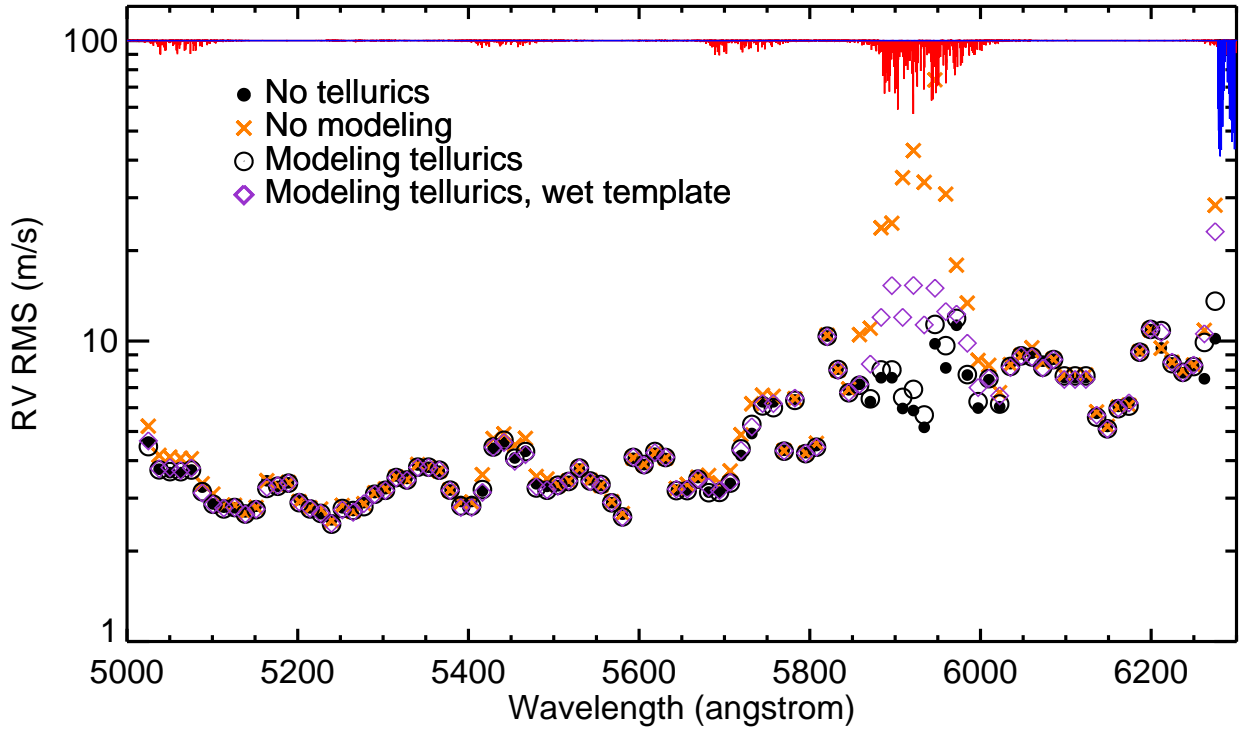} 
\caption{RV RMS for each 50 \AA\ spectral region vs. wavelength in \AA\ for simulated observations of HD 185144 with resolution $R=120$k and SNR = 200. The black dots show the RV precision of telluric-free spectra, showing the photon-limited precision across the iodine region. The orange crosses show the RV precision of simulated spectra with telluric contamination injected but no treatment for the tellurics, using a variation of airmass from real observations and PWV values of the Kitt Peak Observatories in 2016. The hollow circles illustrate how the the precision improves when the telluric absorption lines are being modeled in the Doppler code (but the code uses a DSST with PWV = 1 mm). The hollow diamonds are similar to the hollow circles, but use a DSST with PWV = 5 mm. Water and oxygen absorption lines are shown on top in red and blue, with arbitrary depths.
\label{fig:impact}}
\end{figure*}

Figure~\ref{fig:impact} illustrates the impact of telluric contamination in RV precision as a function of wavelength. The overall increased RV RMS was due to the added RV RMS from the wavelength regions with more water or oxygen absorption, e.g., around 5900 \AA\ and 6300 \AA\ (the orange crosses in Figure~\ref{fig:impact}). Modeling very effectively brings down the RV RMS, making almost full recovery from the damage (i.e., comparing the black circles vs.\ the black dots), unless the stellar template has significant telluric absorption (PWV = 5~mm; purple diamonds). Figure~\ref{fig:impact} is an illustrative case using simulated data on HD 185144 with SNR = 200 and R = 120k, but the general trend and conclusion were the same for simulations with other stars/scenarios.

Overall, tellurics have such a small impact on the RVs primarily because the wavelength regions with relatively heavy telluric contamination were assigned small weights when the RVs from all wavelengths were combined. The telluric-contaminated regions produced RVs with a higher RMS, as shown in Figure~\ref{fig:impact}, so they naturally received low weights, because weights were computed based on the RV RMS of each region over time. This was particularly true for the oxygen region, where deep oxygen lines induced large RV biases as the stellar lines were red-/blue-shifted back and forth across these oxygen lines due to the barycentric motion of the Earth. The entire region with oxygen absorption was actually mostly rejected and not used in the weighted average calculation due to the large intrinsic errors (see the difference between the hollow circles and solid dots in Figure~\ref{fig:impact}). This was still true even when we added telluric modeling due to the oxygen lines in the stellar template.

\begin{figure}
\centering
\includegraphics[scale=0.35]{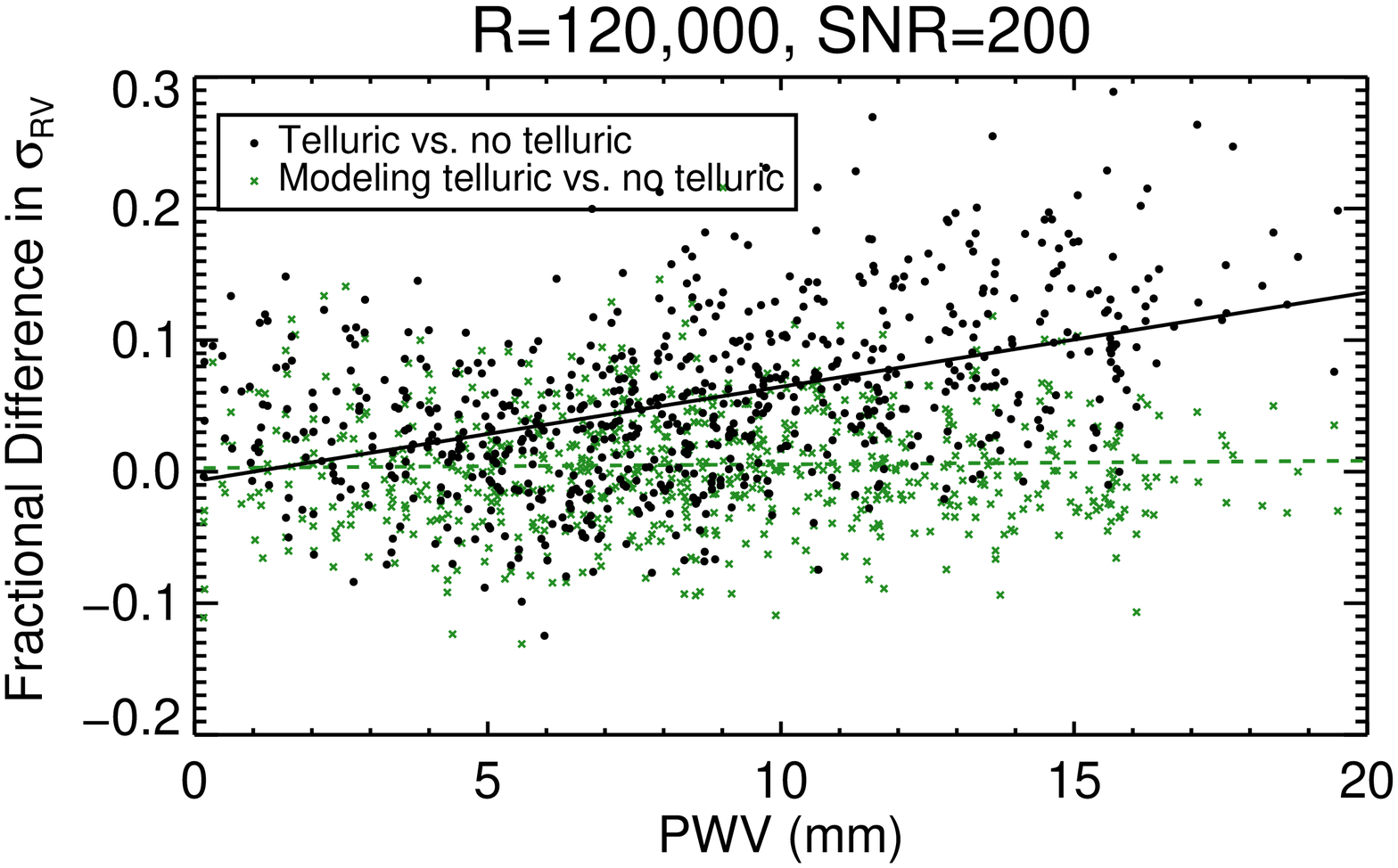} 
\caption{Changes in the internal RV error (``photon-limited'' precision) for each observation due to tellurics vs.\ PWV. Each black dot is the fractional difference in the reported internal RV error ($\sigma_{\rm RV}$) between a telluric-contaminated spectrum, but not being treated for tellurics, and its paired telluric-free spectrum. The case illustrated here is for HD 185144 at R = 120k with SNR = 200. Each green cross is similar, but is the fractional difference between a telluric-contaminated spectrum with forward modeling of telluric lines and its paired telluric-free spectrum. The black line is a linear fit to the black dots, and the green dashed line is a linear fit to the green crosses. See Section~\ref{sec:impact} for more details.
\label{fig:errimpact}}
\end{figure}

Besides adding scatter to the RVs across multiple nights, telluric contamination also degraded the internal RV precision of each nightly observation. Figure~\ref{fig:errimpact} illustrates this effect, where the black dots show the increase in the reported internal RV errors for each observation as the PWV increases. The internal RV error for an observation was derived by evaluating the scatter in the RVs reported by all chunks internally within each observation. This error increased up to $\sim$10\% or so, consistent with the fact that the overall RMS did not increase by more than a few percentage points in the telluric vs.\ telluric-free cases (Table~\ref{tab:simulation}). Such an increase in the internal errors disappeared once we invoked modeling of tellurics, as illustrated in the green crosses. 

There was no significant dependence on the PWVs in terms of the amount of bias induced in the RV of each observation. That is, how much the RV deviated from the true RV (0 m/s) does not appear to strongly depend on the PWVs (tested to be statistically insignificant). This lack of significant dependence on the PWVs can also be seen in Figure~\ref{fig:rvbc}. In fact, the dependence of RV bias on PWVs only started becoming pronounced in test simulations with unrealistically high PWV values (e.g., 10 times the values used here, typically 50~mm or higher).

\subsection{Impact of Telluric Contamination on RVs: Spurious Periodic Signals}\label{sec:impactper}

\begin{figure*}
  \centering
\subfloat{\includegraphics[scale=0.38]{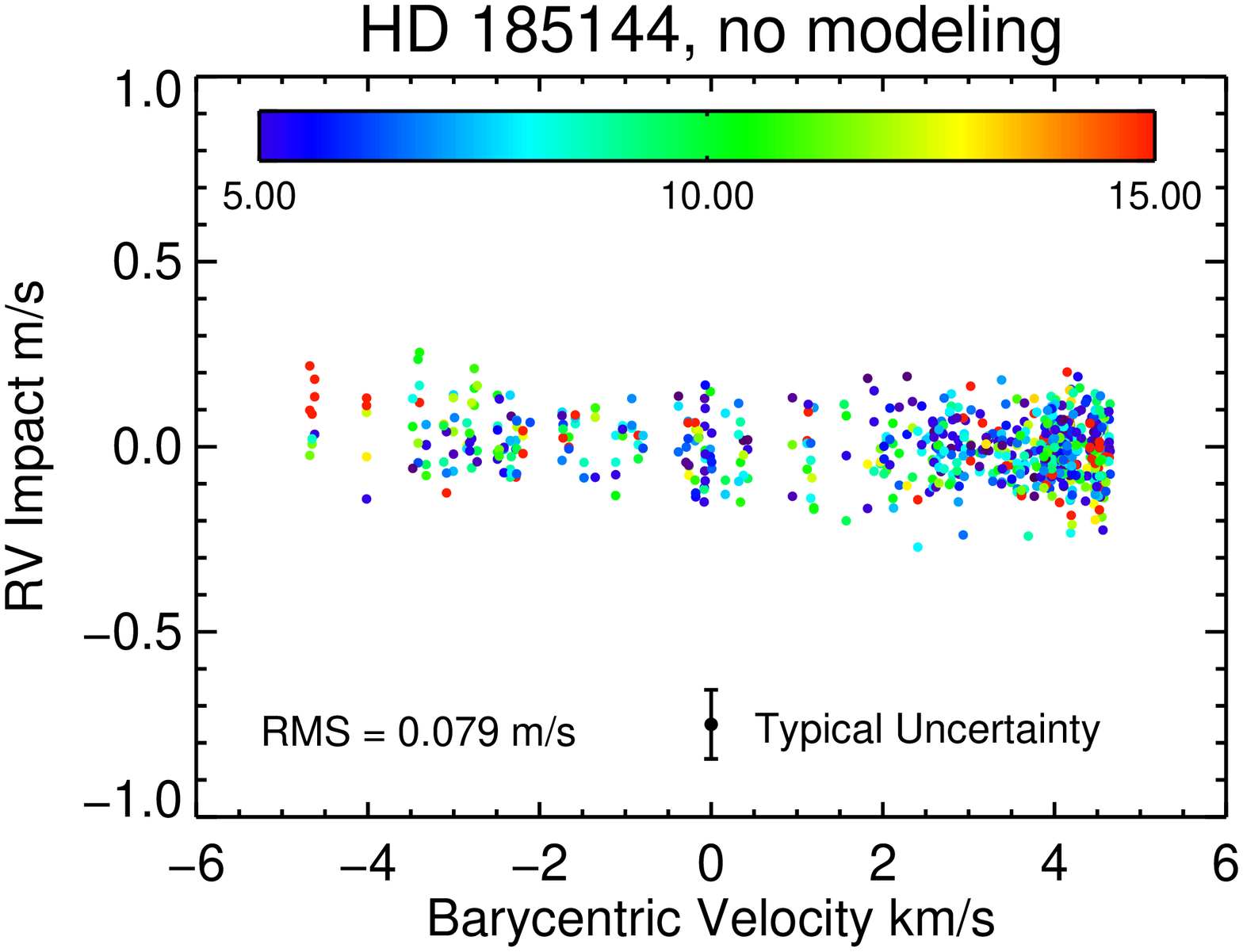}}
\subfloat{\includegraphics[scale=0.38]{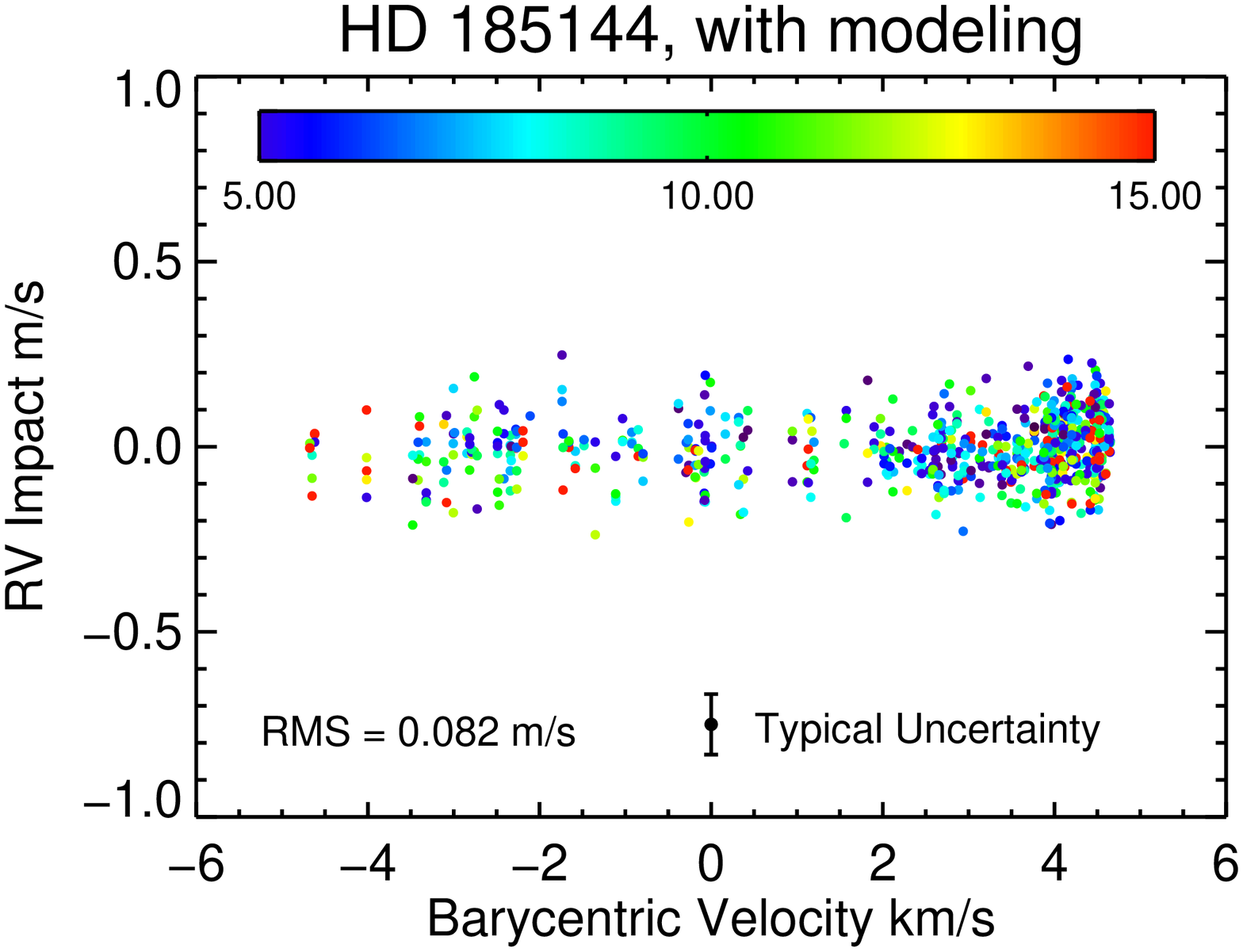}}\
\subfloat{\includegraphics[scale=0.38]{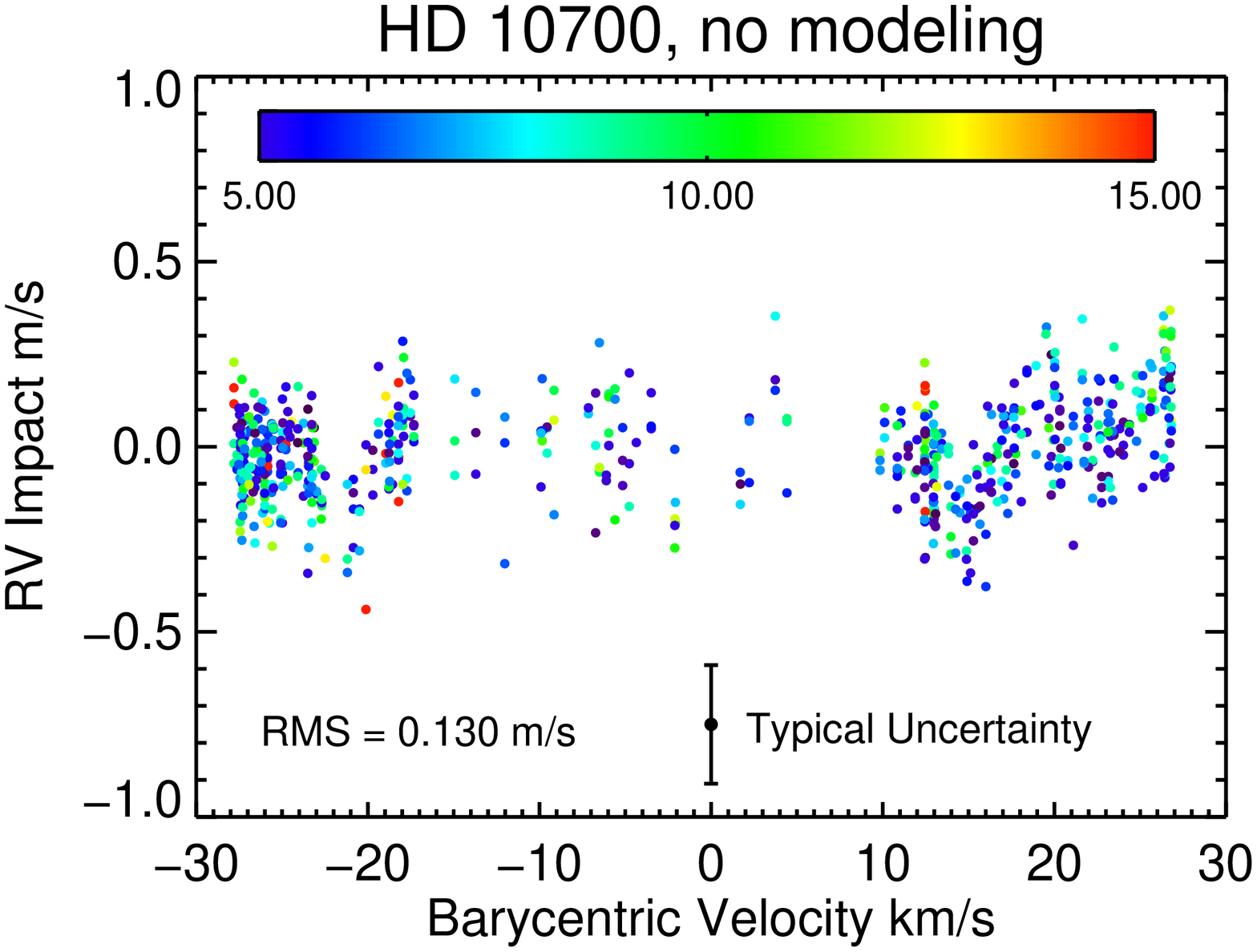}}
\subfloat{\includegraphics[scale=0.38]{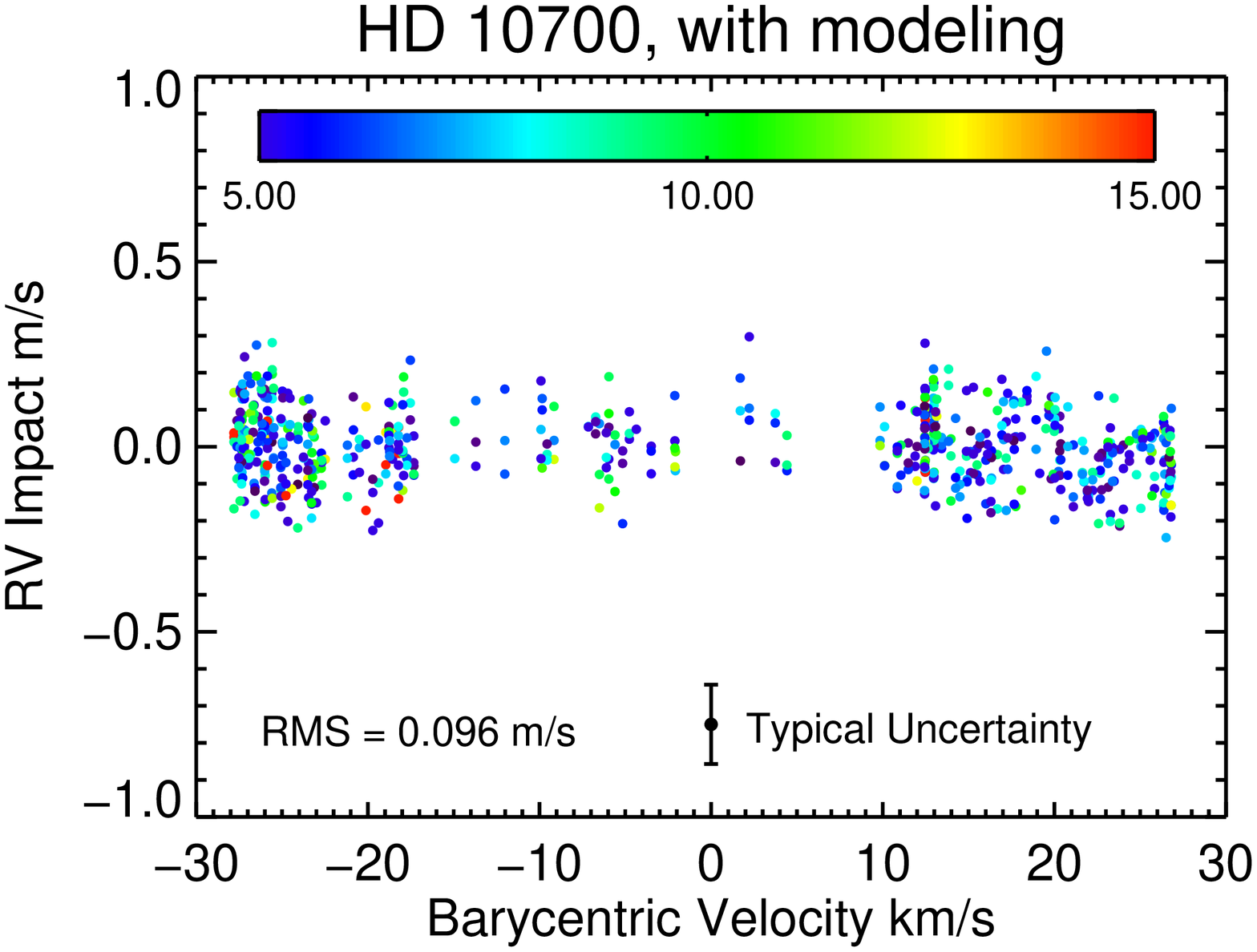}}\
\subfloat{\includegraphics[scale=0.38]{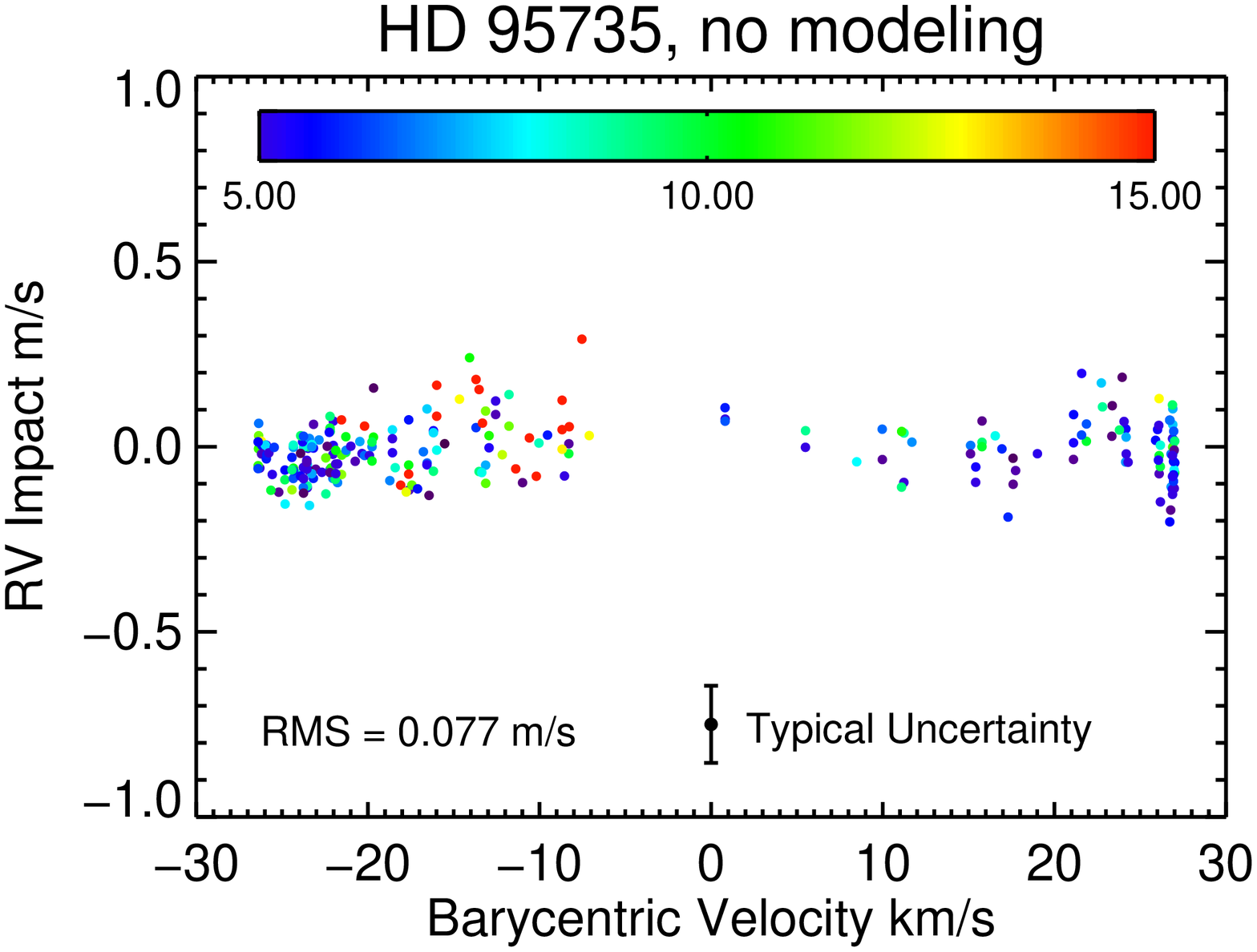}}
\subfloat{\includegraphics[scale=0.38]{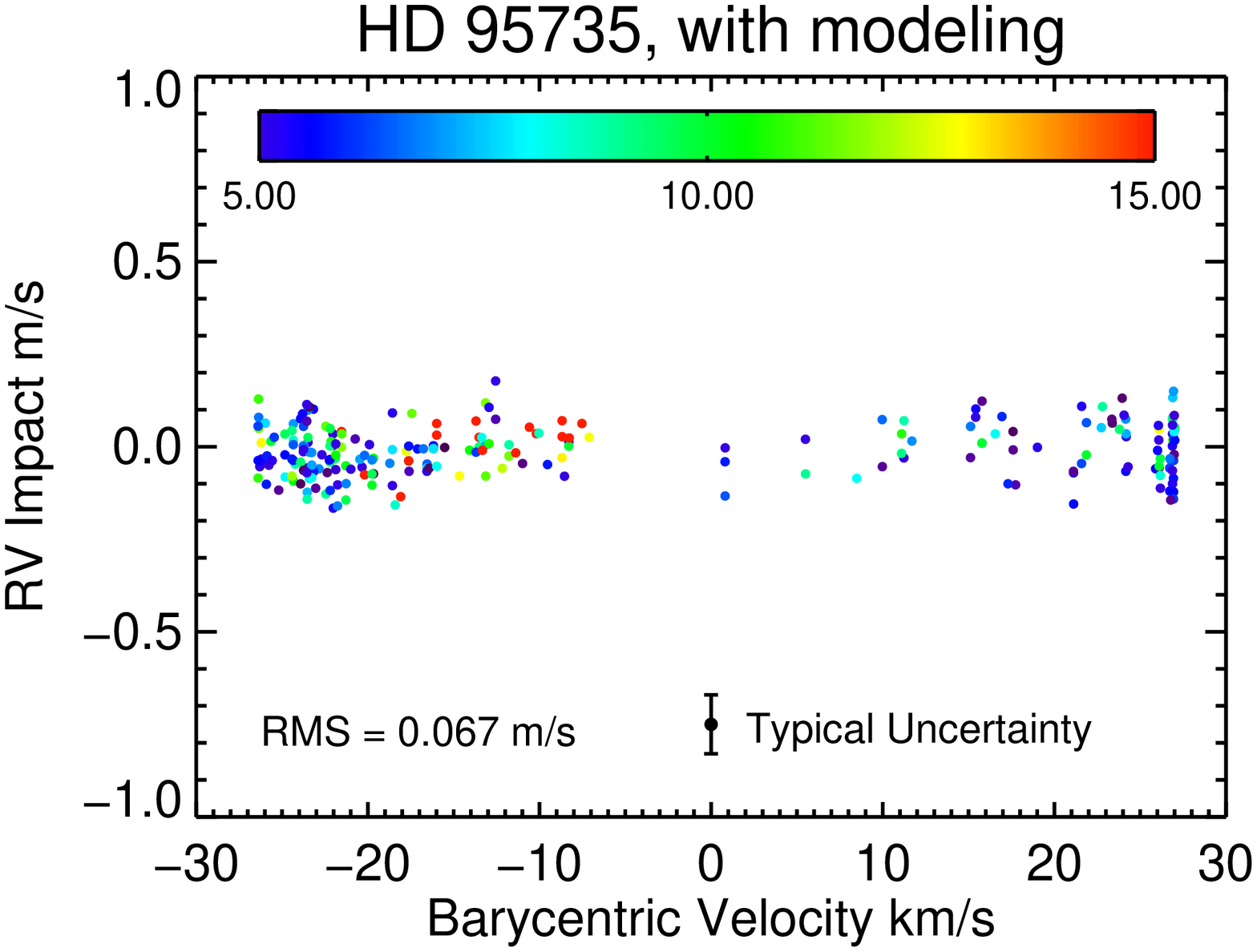}}\
\caption{Difference in RVs between telluric-contaminated and telluric-free spectra vs.\ barycentric velocity of the star for each observation in each star (labeled in subplot titles). The left columns plot the RV difference when forward modeling of tellurics is not invoked, while the right columns show the results with modeling of tellurics. Both sets of simulations used DSST with PWV = 1~mm. Typical internal RV error for each observation is plotted and labeled as "Typical Uncertainty." The color scale denotes the PWV in mm. Modeling is clearly an effective method to remove systematics caused by tellurics. Note that the BC span of HD 185144 is very small. See Section~\ref{sec:impact} for more details.
\label{fig:rvbc}}
\end{figure*}

Telluric contamination could also induce spurious periodic signals in RVs, typically at periods of one year and its harmonics. Such periodicity is due to the barycentric motion of the Earth, which causes the stellar lines to move back and forth on top of the static telluric lines in wavelength space. This back-and-forth motion has a period of one year, and thus, the induced RV bias would also have this signature timescale. Such an ``annual'' systematic effect is best illustrated in RV-BC space, BC being the barycentric velocity of the star. Figure~\ref{fig:rvbc} illustrates this systematic error using simulations with noise-free spectra (because the effects would be buried in noise when photon noise is added). The left panels are for RVs extracted without modeling the tellurics, showing correlation or pattern in the RV-BC plane. The right panels illustrate that the RV-BC correlation is reduced with modeling of the tellurics.

\begin{figure*}
  \centering
\subfloat{\includegraphics[scale=0.55]{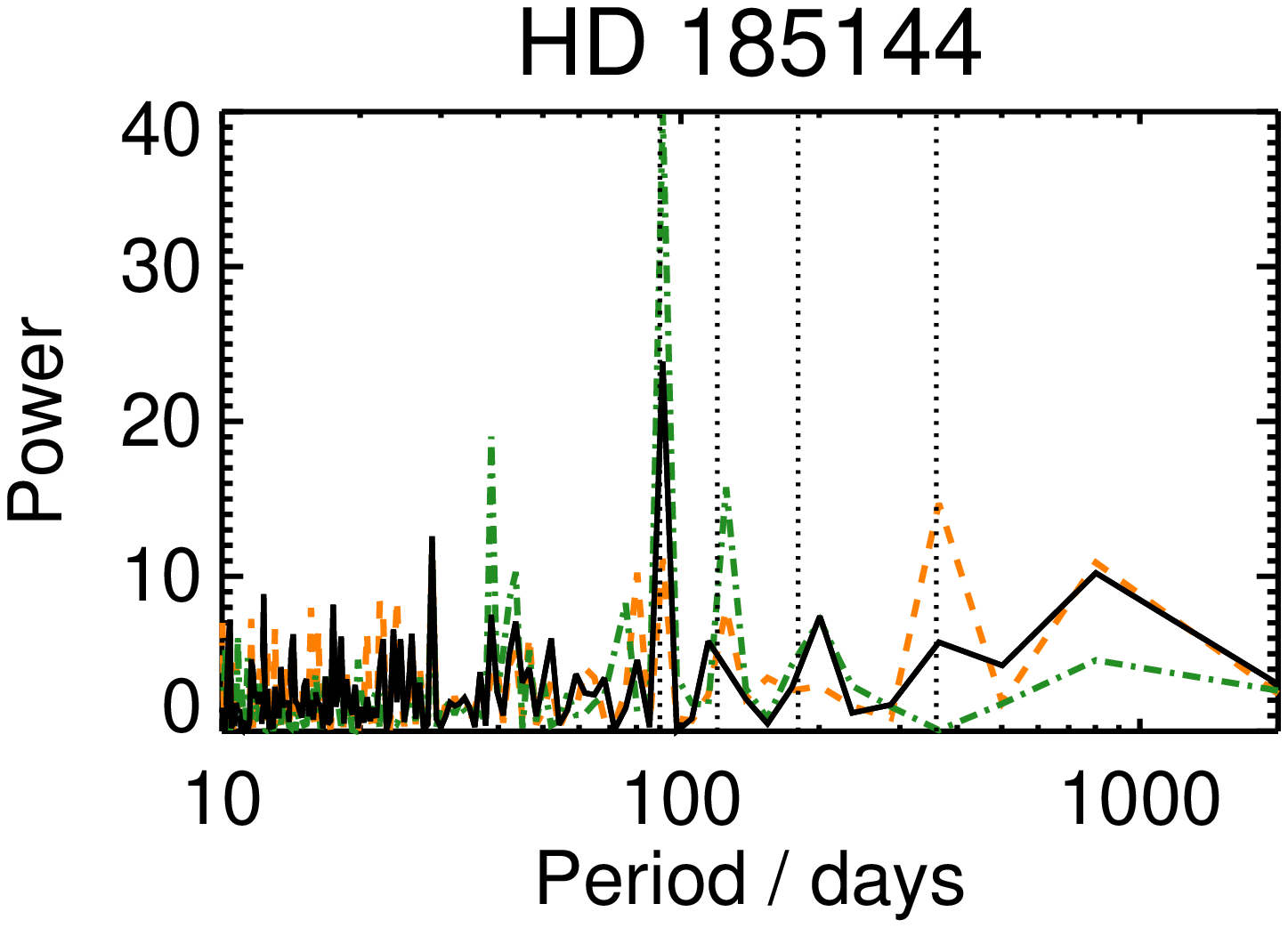}}
\subfloat{\includegraphics[scale=0.55]{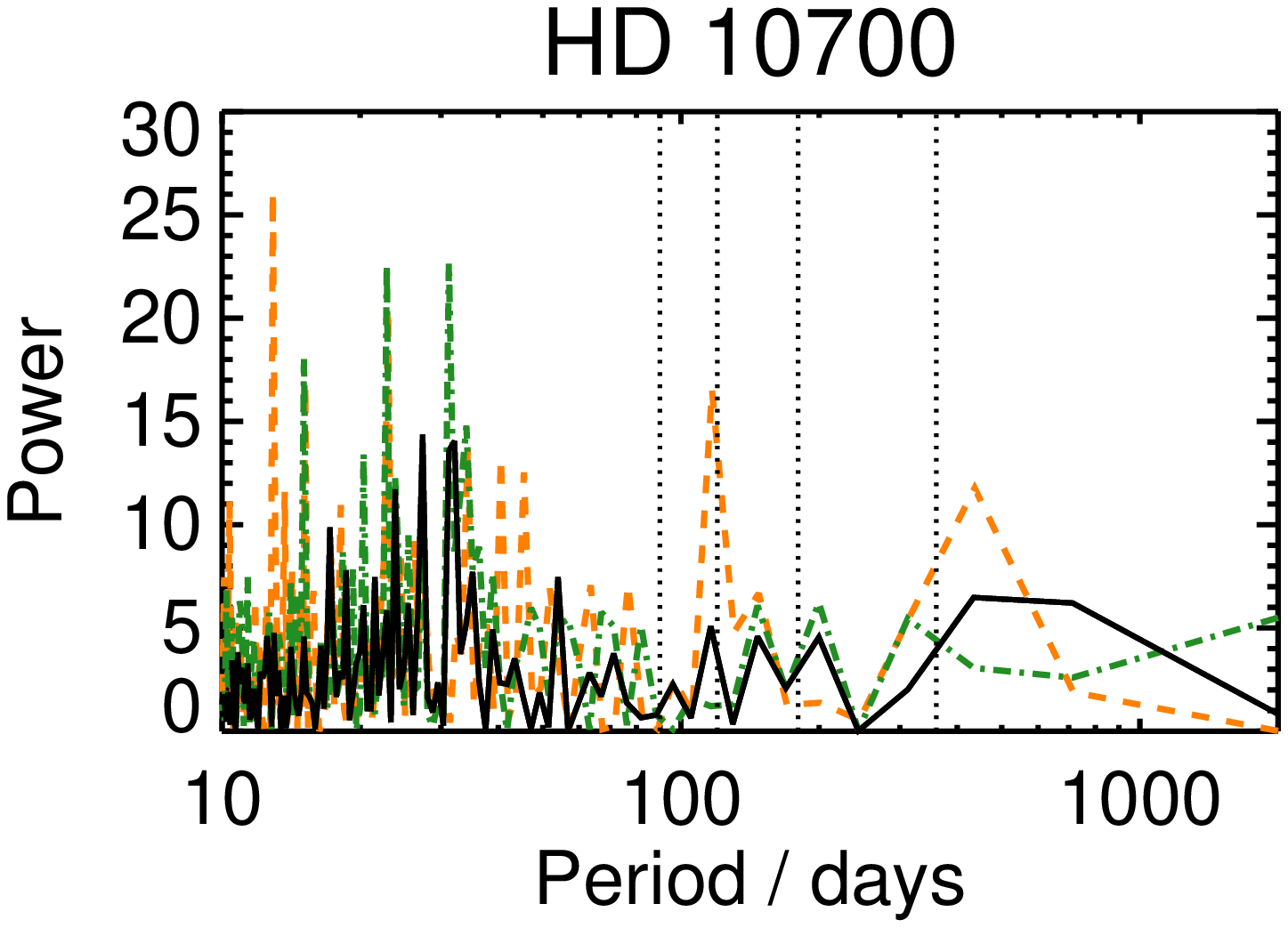}}\\
\subfloat{\includegraphics[scale=0.55]{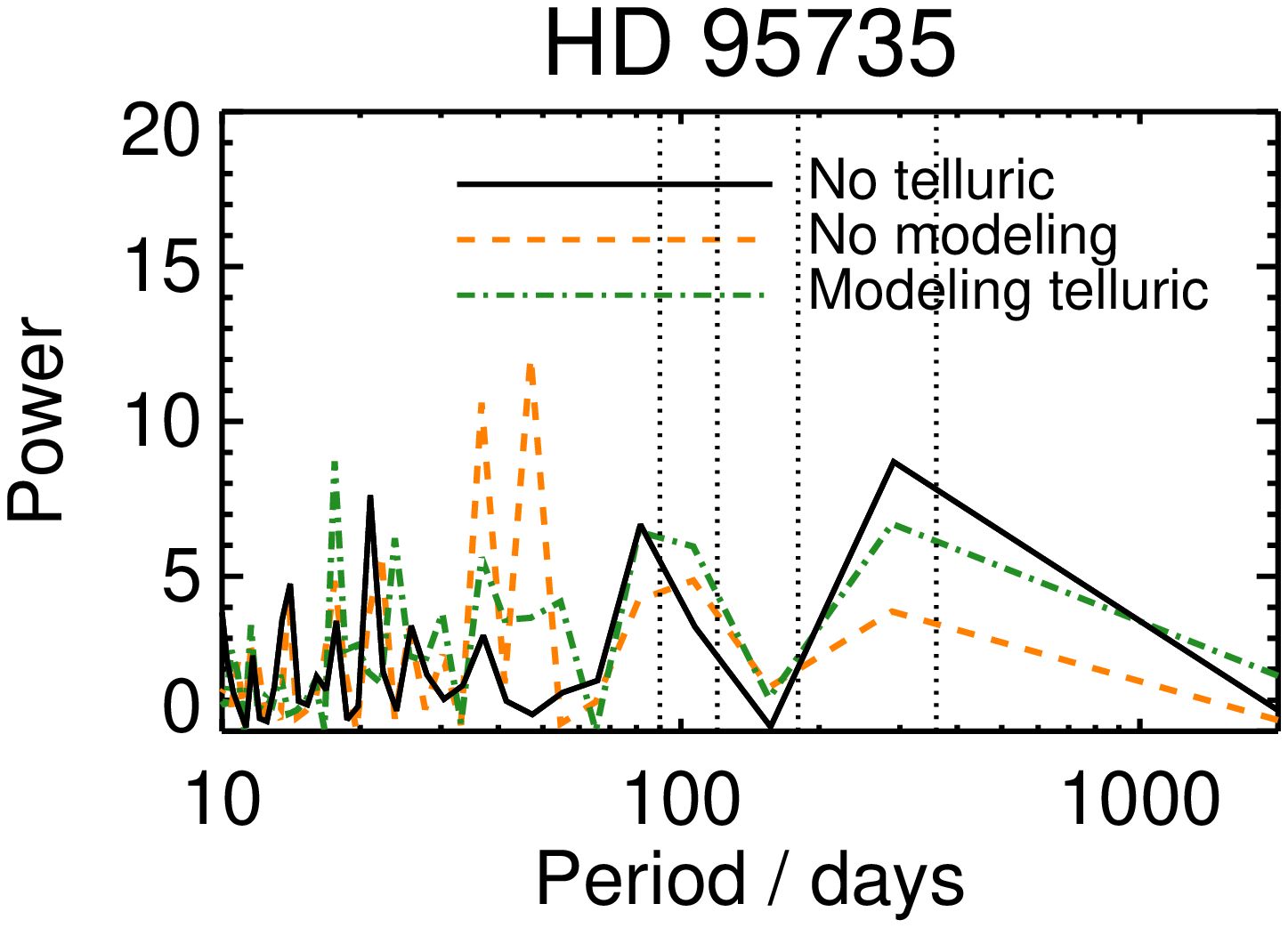}}
\caption{Periodograms of RVs from noise-free spectra at $R = 120$k from simulated spectra without tellurics (solid black), telluric-contaminated spectra with no treatment (no modeling; dashed orange), and telluric-contaminated spectra with forward modeling of tellurics (green dot dashed). The vertical lines are at 90, 180, and 360 days (one year and harmonics), where systematics due to spectra contamination are most likely to occur. Telluric contamination introduces some aliasing in periodograms but not a significant amount. See Section~\ref{sec:impact} for more details.
\label{fig:scargle}}
\end{figure*}

Any strong correlation in RV and BC would translate into periodic signals with periods of one year and/or its harmonics. Figure~\ref{fig:scargle} shows the Lomb-Scargle periodograms for our three test stars for RVs extracted from noise-free spectra without tellurics (black solid), with tellurics added but no modeling of the tellurics (orange dashed), and with tellurics and with modeling (green dash-dotted). The black solid lines are illustrating the window function and any periodic signals induced by systematics intrinsic to the RV extraction pipeline. Comparison between the orange dashed line and the black solid lines reveals the spurious periodic signals introduced by the telluric contamination, and they show up near 360 days, 120 days, 60 days, and so on, as expected. The green dash-dotted lines, i.e., modeling the tellurics, basically recover the black lines, with the exception of HD 185144. For HD 185144, because of its small BC span, the telluric lines in its stellar template are never completely unblended from the telluric lines in the observed spectra being modeled, and therefore, modeling of the tellurics has the least restoration power.

As shown in Figure~\ref{fig:rvbc}, the amplitude of the added spurious signal is on the order of $\sim$20 cm/s, which is very small given the typical photon-limited RV precision of iodine-calibrated spectra. Therefore, in typical real-life scenarios, telluric contamination does not pose a concern in terms of adding spurious signals for iodine-calibrated RV spectrographs aiming at $\sim$1--2 m/s precision.

\subsection{Retrieval of Precipitable Water Vapor via Modeling}\label{sec:retrieval}

The best-fit PWV values are a natural product of our forward modeling process. Since our code assumed no prior knowledge on PWV, how well we recovered the PWV values vs.\ the real input values would provide a gauge on the quality of retrieved PWVs for real observations. Retrieving the PWVs in the iodine region could be very useful, as it is one of the relatively clear regions in terms of tellurics (without heavy blending of stellar lines, \rev{especially for G dwarfs,} and telluric lines of multiple molecular species). For example, the PWV values derived from this wavelength region can then be used for modeling the spectrum in the NIR, where telluric lines are deeper and line blending is more severe, posing more challenges to PWV retrievals.

\begin{figure}
\includegraphics[scale=0.4]{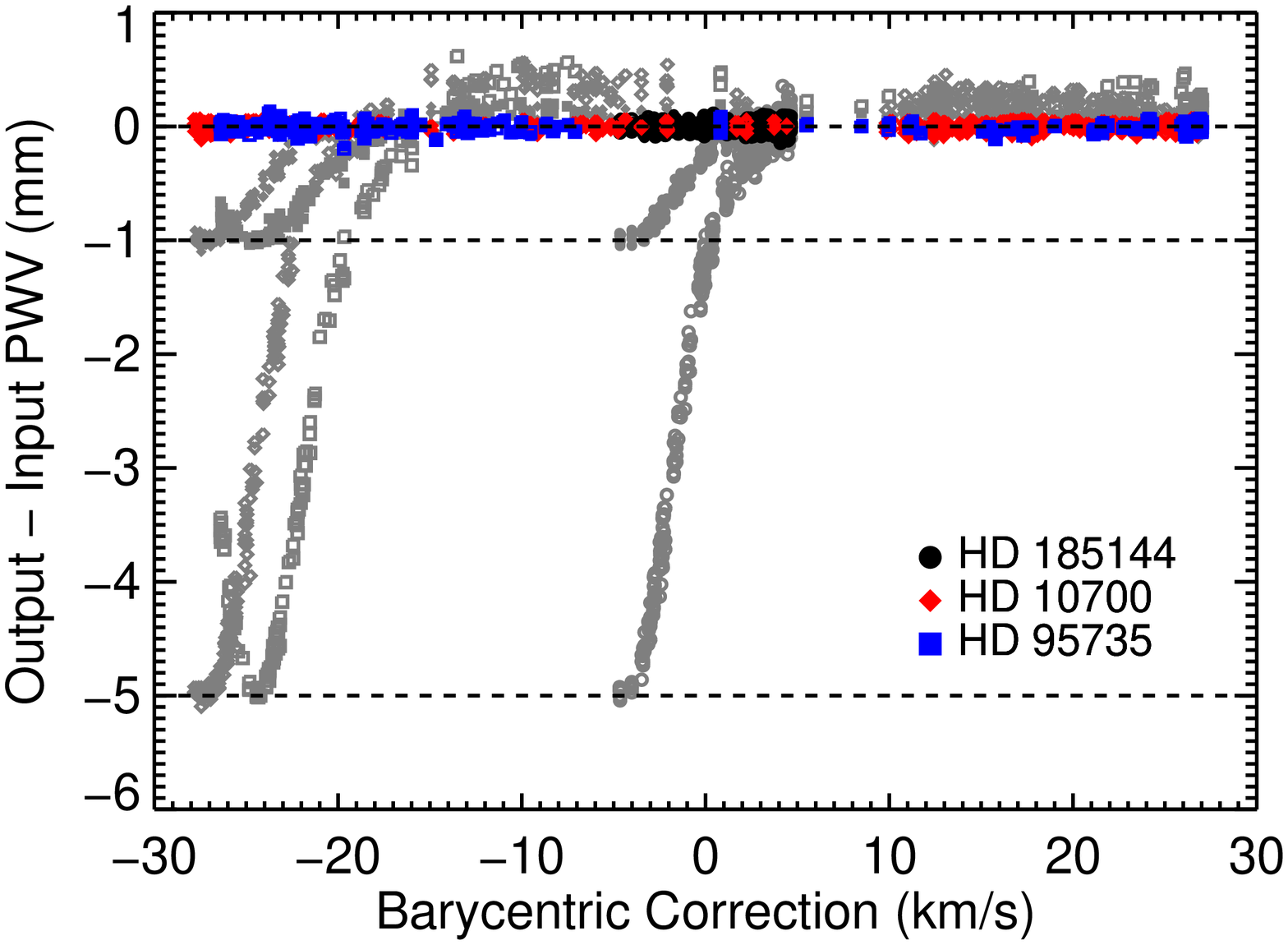} 
\caption{Difference in input and output (best-fit) PWV values vs.\ barycentric velocity (or barycentric correction, BC) of the star for each observation. The solid symbols are for simulations with DSST having PWV = 1 mm, and the hollow symbols are for DSST with PWV = 5 mm. All spectra have $R=120$k and SNR = 200. \rev{This illustrates the excellent agreement between the input and retrieved PWV values, with an RMS of the residuals being 0.04~mm. The gray symbols are for simulations with DSSTs where the water lines have not been removed, and it illustrates the bias in PWV retrieval caused by tellurics in the DSSTs. } 
The bias reaches its maximum when the BC of the star at the observation is the same as the BC of the DSST observation, i.e., when the tellurics in the observation and the DSST are at the same wavelengths. See Section~\ref{sec:retrieval} for more.
\label{fig:pwvbc}}
\end{figure}

When using a telluric-free stellar template in the forward modeling, PWV values can be recovered to relatively high accuracy, typically to better than 0.5~mm. This is on par or better than other PWV monitoring methods using NIR photometry \citep{baker2017} or GPS \citep{blake2011}. However, as mentioned in Section~\ref{sec:simspec}, it is unrealistic to have a telluric-free stellar template. As a result, the presence of telluric lines in the stellar template will bias the PWV measurements, especially when the telluric lines in the template are blended with the ones in the observed spectra being modeled. This bias is illustrated \rev{by the gray symbols} in Figure~\ref{fig:pwvbc}. When a star has a small BC span, such as HD 185144 (black points), this bias affects more observations, because the relative redshift between template and observations are always small. The amount of bias depends on the depths of water lines in the stellar template (hollow symbols vs.\ filled symbols).

\rev{We have also divided out the telluric lines in the simulated DSSTs in order to achieve better accuracy in the retrieved PWV values. This is trivial for the simulations, as we know exactly the input PWV, and the wavelength solution for the simulated DSSTs is also perfect so that we can divide out the telluric lines ``perfectly".\footnote{In reality, one can never divide out telluric lines perfectly, because the observed spectrum was convolved with the spectral PSF, and the mathmatically, the telluric absorption was multiplied on top of the stellar absorption lines before this convolution took place. Since convolution does not distribute over multiplication, dividing the telluric lines in the convolved spectrum is mathmatically incorrect.} As shown in Figure~\ref{fig:pwvbc}, the retrieved PWV values using a cleaned DSST match very well with the input PWV values. For simplicity, we did not list the RV RMS values from analyses using these cleaned DSSTs, as they are essentially the same (within 1~cm/s or the numerical precision of our Doppler code) as the RV RMS from the telluric free simulations. Such a complete elimination of the effects of tellurics, however, is unfortunately not true for real observations. As listed in the last row of Table~\ref{tab:simulation}, RV analyses on real observations using ``cleaned DSSTs" did not return better RV RMS consistently (see Section~\ref{sec:realobs} for more). Therefore, it is unclear whether cleaning the telluric lines in DSSTs would result in any significant benefit besides accuracy in the retrieved PWV for typical iodine-calibrated RV instrument with 1--2~m/s precision.}


\section{Fitting Tellurics in Real Observations}\label{sec:realobs}

As demonstrated in the sections above, telluric contamination does not add a significant amount of RV errors or biases to the iodine-calibrated data, especially in consideration of the typical SNR and precision. To confirm this result, we added modeling of tellurics when extracting RVs from real \keck\ data, following the procedure described in Section~\ref{sec:dopcode}. The results are listed in the bottom two rows of Table~\ref{tab:simulation}. As expected, there was virtually no effect in terms of RV RMS after adding in telluric modeling, because of the SNR and the fact that the \keck\ RV systematics was probably dominated by other sources instead of tellurics (see \citealt{thesis} for more). Also, as expected, adding telluric modeling has no significant effects in terms of changes in the systematics in the RV-BC plane or added spurious signals in the periodograms.
\begin{figure}
\includegraphics[scale=0.4]{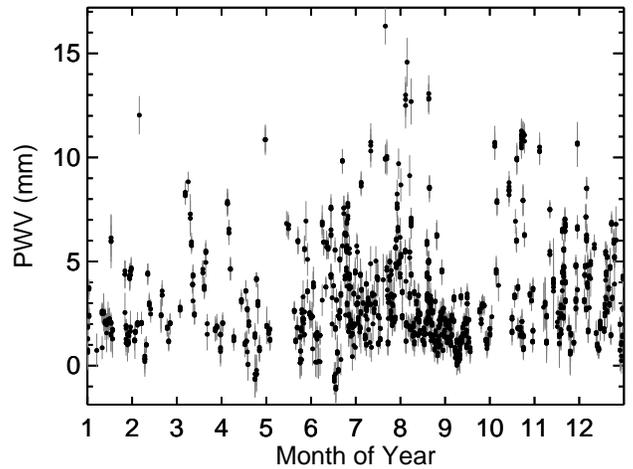} 
\caption{PWV values derived from real \keck\ observations of HD 185144, HD 10700, and HD 95735 plotted against month of year. The seasonal variation and absolute PWV values are consistent with those reported in the literature \citep[e.g.,][]{garcia2010}. \rev{The PWV values are derived from RV analyses using ``cleaned DSSTs" where water lines have been divided out. The error bars are from the scatter in the PWV values derived among different spectral chunks within each observation.} See Section~\ref{sec:realobs} for more details.
\label{fig:pwvrealobs}}
\end{figure}

Figure~\ref{fig:pwvrealobs} shows the retrieved PWV values from real \keck\ observations on our three stars. The overall trends and range of values of the PWV were consistent with other site measurement for Mauna Kea, such as \cite{garcia2010}. \rev{To avoid biases in the retrieved PWV values caused by water lines in the DSST (Figure~\ref{fig:pwvbc}), we cleaned up the DSSTs beforehand to eliminate the water lines in them as much as possible. We visually examined the DSST spectral chunks with relatively deep water lines, and adjusted the depth and width (i.e., resolution) of the water lines to match with the observed (but deconvolved) lines in the DSST as much as possible. We then divided out the water lines given the visual best-fit width and depth (PWV) across the entire DSST to arrive a cleaned DSST. There are still some negative PWV values, which are probably a result of residuals from this division, i.e., an underestimated PWV. Unfortuantely, as shown in the last row of Table~\ref{tab:simulation}, these cleaned DSSTs did not bring consistent improvements on the RV precision in real observations, which could be due to incomplete removal of water lines and/or additional error induced in the division process.}

\section{Conclusion and Future Work}\label{sec:summary}

Using simulated spectra similar to \keck's and real weather data from Kitt Peak, we have characterized the effects of telluric contamination in iodine-calibrated precise RVs. We conclude that telluric contamination introduces additional errors and systematics on the order of 10 cm/s for iodine calibrated RVs, similar to previously reported for HARPS in \cite{cunha2014} and \cite{artigau2014}. This amount of error is essentially negligible given the typical SNR and precision achieved by real on-sky iodine-calibrated RV spectra. \rev{For drier sites such as Mauna Kea and the Canary Islands, the effects of tellurics for iodine-calibrated RVs would be even smaller compared to the conclusion from this study based on the weather data of Kitt Peak.}

We found that the adverse effect of telluric contamination is contained to $\sim$10 cm/s, mostly for two reasons: the shallow lines in the iodine region (500-620~nm) and the down weighting of the telluric-contaminated spectral region due to the low RV precision caused by tellurics. 

We also added telluric lines as an additional component in the forward modeling process when extracting RVs from iodine-calibrated data, and we conclude that modeling effectively mitigates the effects of telluric contamination, both in terms of minimizing the added RV errors and in correcting the spurious periodic signals induced by tellurics. We have demonstrated that modeling the water lines does not require prior knowledge on the PWV values, and that modeling can retrieve the PWV information somewhat accurately. However, one should caution that, when the BCs of the RV observations are close to the BC of the stellar template used in the forward modeling process, there will be significant bias in the retrieved PWV values due to telluric line blending. Unfortunately, telluric lines in the stellar template observations are very hard to remove without distorting the template and creating a net decrease in RV precision. It is possible that template-free methods for extracting RVs (e.g., \citealt{gao2016}, \citealt{czekala2017}) would be free of the problems caused by telluric lines in the stellar templates, which are generally derived from observations.

In practice, ignoring tellurics and not modeling them will not compromise the RV precision in iodine-calibrated RV data in any significant way, as long as all spectral regions are carefully evaluated and weighted before being combined to derive a final RV. However, for any RV program using iodine that aims at $<$~1~m/s precision, we recommend that telluric modeling be added to the forward modeling process, especially if the humidity of the telescope site is similar to or higher than Kitt Peak (with a median PWV of $\sim$5~mm). We also highly recommend that stellar template observations be taken on relatively dry nights (ideally PWV$\sim$5~mm or smaller).

\rev{We notice that spectral resolution does not seem to be critical for the resilience against or the ability to recover from the impact of tellurics for iodine-calibrated RVs, which is perhaps not surprising given that both stellar lines and water lines are basically resolved with R$\sim$70k within the iodine wavelength region. As shown in Table~\ref{tab:simulation}, the results for the R = 120k simulations do not appear to suffer less from tellurics, nor do they recover better when fitting tellurics in comparison to the R = 70k simulations. As the typical iodine-calibrated instruments typically only reach to up to R $\sim$ 130k (e.g., the Planet Finder Spectrograph, PFS, on Magellan, \citealt{crane2010}), we did not perform simulations with an even higher spectral resolution and whether that would bring better performance under the influence of tellurics. However, for the next-generation spectrographs, this could be vitally important, especially for the NIR where telluric lines are copious and often heavily blended with stellar lines, especially for M dwarfs.}

\rev{Importantly though irrelevant to telluric contamination, our simulations, especially the results in Figure~\ref{fig:snrprecision}, demonstrate that the choice of resolution can be important for achieving high RV precision with iodine-calibrated instruments. In particular, for slit-fed instruments or fiber-fed instruments that are resilient against seeing loss, choosing a higher resolution setting (e.g., R = 120k vs.~R = 70k) could boost the RV precision considerably given the same exposure time. This boost of RV precision has been verified in the on-going surveys using Magellan/PFS, including the RV follow-up on small transiting exoplanets discovered by TESS \citep{ricker2015}. After installing a larger-format CCD with smaller pixels, Magellan/PFS switched its nominal operation mode from R = 80k to R = 130k (from using the $0.5\arcsec$ slit to using the $0.3\arcsec$ slit, or a boost in resolution and RV precision for the same SNR at a factor of 5/3). The photon loss due to a narrower slit was smaller than a factor of $(1-3/5)=40$\% given the superb seeing at Las Campanas Observatory, and as a result, the RV precision has increased considerably without requiring a significant increase on exposure time (e.g., see the difference in old and new PFS data in \citealt{dragomir2019}).\footnote{\rev{Assuming a perfect Gaussian PSF, a 0.3$\arcsec$ slit would lose 32\% of light compared with a 0.5$\arcsec$ slit for a seeing of 0.5$\arcsec$, typical at Las Campanas. Considering the boost in RV precision due to a higher resolution by a factor of 5/3, the net gain in photon efficiency would be 5/3/0.32 = 13\%. Given that total time required to robustly measure a given RV signal is proportional to the square of single-exposure RV precision (e.g., \citealt{howard2016}), this would boost the RV survey efficiency by 30\% (or 15\% if assuming a 0.7$\arcsec$ seeing using a similar argumetn).}} We thus recommend iodine-calibrated RV programs that require high efficiency and high RV precision, such as TESS follow-up programs, adopt a high-resolution mode if possible. Again, for particularly challenging targets with small RV amplitudes ($\sim$1~m/s), we recommend incorporation of telluric contamination to eliminate potential additional error sources and aliases in the frequency domain as much as possible.}

\rev{Although we argue that mitigating tellurics is not vital (though could be important) for iodine-calibrated RVs, for the next-generation RV spectrographs that aim at a precision of 10 cm/s, the errors induced by telluric contamination definitely cannot be ignored.} As the amplitude of the additional errors and spurious signals added by tellurics is essentially a result of the competition between the Doppler content in the stellar spectrum vs.\ the telluric spectrum, it is reasonable to speculate that future spectrographs with higher precision but extending to wider and redder wavelength ranges will face a larger amount of systematics, as the telluric absorption increases towards the red. \rev{How to mitigate telluric contamination effectively across a broad range of wavelength coverage and spectrograph setups is the topic of many on-going and future studies (e.g., \citealt{plavchan2018}).\footnote{\rev{A full report of the \textit{EarthFinder} Probe Mission Concept Study can be found at \url{https://smd-prod.s3.amazonaws.com/science-red/s3fs-public/atoms/files/Earth_Finder_Study_Rpt.pdf}, and Section 1.3 focuses on mitigating telluric contamination.}}}



\acknowledgements

We thank John A. Johnson for providing a copy of his Doppler code and his help with incorporating the code. The authors also thank Debra Fischer for her assistance in this regard and Dr.\ Paula Coelho for providing the high resolution, high sampling M dwarf synthetic spectra. Finally, we thank Drs.\ Eric Ford, Suvrath Mahadevan, Jim Kasting, and Larry Ramsey for their input on this work.

S.X.W.\ acknowledges support from NASA Earth and Space Science Graduate Fellowship (2014-2016). J.T.W.\ and S.X.W.\ acknowledge support from NSF AST-1211441. This work was also partially supported by funding from the Center for Exoplanets and Habitable Worlds, which is supported by the Pennsylvania State University, the Eberly College of Science, and the Pennsylvania Space Grant Consortium. 

We appreciate the work done by the observers who took the data using \keck\ on HD 185144, HD 10700, and HD 95735, which has enabled this work.

The work herein is based on observations obtained at the W. M. Keck Observatory, which is operated jointly by the University of California and the California Institute of Technology.  The Keck Observatory was made possible by the generous financial support of the W.M. Keck Foundation.  We wish to recognize and acknowledge the very significant cultural role and reverence that the summit of Mauna Kea has always had within the indigenous Hawaiian community.  We are most fortunate to have the opportunity to conduct observations from this mountain.

This work has made use of NASA's Astrophysics Data System Bibliographic Services. This research has made use of the SIMBAD database, operated at CDS, Strasbourg, France \citep{simbad}.

\software{TERRASPEC \citep{Bender2012}, TAPAS \citep{tapas}, PHOENIX (v15.5; \citealt{hauschildt1993,hauschildt2006,baron2007}}.

\end{CJK*}

\bibliography{references}

\end{document}